\documentclass[conference,compsoc]{IEEEtran}

\usepackage{hyperref}
\usepackage{graphicx}
\usepackage{multirow}
\usepackage{float}
\usepackage{amsmath}
\usepackage{amssymb}
\usepackage{algorithm}
\usepackage{algorithmic}
\usepackage{enumitem}
\usepackage{framed}
\usepackage{subcaption}
\usepackage{placeins}
\usepackage[table,xcdraw]{xcolor}

\DeclareMathOperator*{\argmin}{arg\,min}

\ifCLASSOPTIONcompsoc
  \usepackage[nocompress]{cite}
\else
  \usepackage{cite}
\fi

\ifCLASSINFOpdf
\else
\fi

\begin{document}
\pagestyle{plain}

\title{\large Practical, Generalizable and Robust Backdoor Attacks on Text-to-Image Diffusion Models}

\author{\IEEEauthorblockN{Haoran Dai\IEEEauthorrefmark{2},
Jiawen Wang\IEEEauthorrefmark{2},
Ruo Yang\IEEEauthorrefmark{3},
Manali Sharma\IEEEauthorrefmark{3},
Zhonghao Liao\IEEEauthorrefmark{4},
Yuan Hong\IEEEauthorrefmark{5}, and
Binghui Wang\IEEEauthorrefmark{2}}
\IEEEauthorblockA{\IEEEauthorrefmark{2}Illinois Institute of Technology.
\{hdai10, jwang306\}@hawk.iit.edu, bwang70@iit.edu}
\IEEEauthorblockA{\IEEEauthorrefmark{3}Samsung. 
r.yang@partner.samsung.com, manali.s@samsung.com}
\IEEEauthorblockA{\IEEEauthorrefmark{4}Milwaukee School of Engineering. 
liao@msoe.edu}
\IEEEauthorblockA{\IEEEauthorrefmark{5}University of Connecticut.
yuan.hong@uconn.edu}
}

\maketitle

\begin{abstract}
Text-to-image diffusion models (T2I DMs) have achieved remarkable success in generating high-quality and diverse images from text prompts, yet recent studies have revealed their vulnerability to backdoor attacks. Existing attack methods suffer from critical limitations: 1) they rely on unnatural adversarial  prompts that lack human readability and require massive poisoned data; 2) their effectiveness is typically restricted to specific models, lacking generalizability; and 3) they can be mitigated by recent backdoor defenses.

To overcome these challenges, we propose a novel backdoor attack framework that achieves three key properties:
1) \emph{Practicality}: Our attack  requires only a few stealthy backdoor samples to generate arbitrary attacker-chosen target images, as well as ensuring high-quality image generation in benign scenarios.
2) \emph{Generalizability:} The attack is applicable across multiple T2I DMs without requiring model-specific redesign.
3) \emph{Robustness:} The attack remains effective against existing backdoor defenses and adaptive defenses.
Our extensive experimental results on multiple T2I DMs demonstrate that with only 10 carefully crafted backdoored samples, our attack method achieves $>$90\% attack success rate with negligible degradation in benign image generation quality. 
We also conduct human evaluation to validate our attack effectiveness. Furthermore, recent backdoor detection and mitigation methods, as well as adaptive defense tailored to our attack are not sufficiently effective, highlighting the pressing need for more robust defense mechanisms against the proposed attack.

\end{abstract}

\IEEEpeerreviewmaketitle

\section{Introduction}

Text-to-Image (T2I) synthesis enables the generation of high-quality images from natural language descriptions. This capability is largely driven by recent developments in diffusion models (DMs) \cite{ho2020denoising,ramesh2022hierarchical,yang2023diffusion}, such as Stable Diffusion~\cite{Rombach_2022_CVPR}, SDXL~\cite{podell2023sdxl}, and Imagen~\cite{saharia2022imagen}. T2I DMs use iterative denoising processes to progressively refine generated images conditioned on text prompts. 
Due to their unprecedented image generation performance, T2I DMs have been adopted in a wide range of domains, including healthcare~\cite{senft2025does,kidder2024advanced}, education~\cite{ali2024picture}, and copyright-sensitive content creation~\cite{zhu2024generative}.
 
Nevertheless, the rapid development and deployment of T2I DMs have often outpaced thorough evaluation of their security properties. 
This oversight is particularly concerning as they are increasingly adopted in high-stakes domains. 
For instance, medical diagnostic systems integrated with T2I DM \cite{kidder2024advanced}, 
once backdoored, could generate misleading medical images that may insert signs of a disease or omit critical anomalies in response to benign prompts, potentially leading to misdiagnosis and compromised patient safety. 
Similarly, if a T2I DM intended for educational use by children is manipulated to generate violent content, it could raise serious psychological and ethical concerns.

Several recent studies ~\cite{chen2023trojdiff,DBLP:journals/corr/abs-2412-11441, chou2023baddiffusion, zhai2023text,du2023stable, struppek2023rickrolling, chou2023villandiffusion, huang2024personalization, vice2024bagm, shan2024nightshade} 
have revealed that T2I DMs are vulnerable to backdoor attacks, primarily due to their dependence on large-scale, publicly available training datasets. These datasets are often uncurated and thus susceptible to data poisoning, wherein an adversary can inject malicious examples to embed specific behaviors into the model. Once trained, the compromised model may produce intentionally incorrect or harmful outputs when triggered by designated inputs. As such, understanding and mitigating the security vulnerabilities of T2I DMs—particularly against backdoor attacks—is a critical and timely research problem in the broader context of AI safety and robustness.

However, current backdoor attack strategies on T2I DMs exhibit several major limitations (More details are deferred to Related Work in Section~\ref{sec:related}):

 \vspace{+0.05in}
\noindent \textbf{1. Unnatural Poisoned Prompts.} 
    Many existing methods \cite{chen2023trojdiff,chou2023baddiffusion,struppek2023rickrolling} use nonsensical or visually unreadable poisoned prompts.
    For example, \cite{struppek2023rickrolling} replaces Latin characters with visually similar but distinct Unicode characters, e.g., the letter “o” in a prompt such as “A boat on a lake” is replaced with the Cyrillic character “\b{o}”, resulting in the modified prompt “A boat \b{o}n a lake”. 
    While visually subtle, such modifications are unlikely to occur in natural user input and can be easily detected via Unicode character detectors.

    \vspace{+0.05in}
\noindent   \textbf{2. Require Massive Poisoned Data.}
    Achieving reasonable attack effectiveness often requires injecting a substantial amount of poisoned data \cite{chou2023baddiffusion,vice2024bagm}. E.g., \cite{vice2024bagm} needs a poisoning rate of 5\% to obtain an attack success rate of 85\%. 
    Yet, despite appearing modest, this rate poses substantial challenges in GenAI---the SOTA DMs are typically trained on datasets containing millions or billions of samples~\cite{rombach2022high}.

\vspace{+0.05in}
\noindent  \textbf{3. Lack of Generalizability.}
   Their attack effectiveness highly depends on specific model architectures. E.g., 
   \cite{huang2024personalization} uses personalization techniques, such as Textual Inversion \cite{wu2023backdooring} and DreamBooth \cite{ruiz2023dreambooth}, to embed backdoors. However, these techniques are tightly coupled with particular model designs, which limits their generalizability to other T2I DMs that may use different architectures or training paradigms.

   \vspace{+0.05in}
\noindent  \textbf{4. Vulnerable to Emerging Defenses.}
    Many  defense methods \cite{sui2024disdetexploringdetectabilitybackdoor,mo2024terd,wang2024t2ishield,chew2024defending} have been developed recently to defend against backdoor attacks on T2I DMs.
    For example, T2IShield~\cite{wang2024t2ishield}, a backdoor detection and mitigation method, shows effectiveness against recent attacks \cite{struppek2023rickrolling, chou2023villandiffusion}. 

In this paper, we consider a {targeted} backdoor attack on T2I DMs, where an attacker inserts a backdoor trigger into a text prompt\footnote{We will interchangeably use (poisoned) ``text", ``prompt", and ``text prompt" throughout this paper.}  such that the poisoned text forces the backdoored T2I DM to output the \emph{target image} (e.g., a sexually explicit or violent image, or a medical image omitting
critical anomalies) that is \emph{arbitrarily} chosen by the attacker. 
Especially, we expect the designed attack to address all the above limitations: it is \emph{practical (i.e., invisible trigger, only requires a few poisoned samples, efficient), generalizable (i.e., applicable and effective across diverse T2I DMs), and robust (i.e., detection and mitigation based defenses as well as adaptive defenses are insufficient)}. 
We propose a novel three-stage attack framework to achieve our 
goals.

First, it ensures semantic trigger concealment by embedding malicious behavior into natural language phrases (e.g., “A mouse and a cat”), where the trigger word “cat” appears contextually natural, aligns with user intent, and avoids detection through lexical anomaly. In particular, we leverage prompt engineering techniques to efficiently
construct such invisible triggers and the poisoned prompts. 

Second, it enhances the semantic similarity between the poisoned prompt and the target image. 
This is challenging due to the nature of invisible triggers—they are typically common vocabulary terms that appear frequently in natural prompts, and T2I DMs have already learned strong correlations between these words and their corresponding visual concepts during pretraining. For instance, if “cat” is an invisible trigger word, due to the model’s extensive exposure to the word “cat” and associated cat images, the learned association is deeply embedded. Modifying this strong correlation to associate “cat” with a different target image requires substantial efforts, as it demands overriding existing representations learned from large-scale real-world data. 
Though a few techniques~\cite{huang2024personalization} attempt to manipulate specific models—e.g., forcing a model to \emph{forget} the learned association between the word “cat” and corresponding images, these approaches rely on architecture-specific modules and are sample and training inefficient.
To address this challenge and consider the attack generality, we adopt CLIP \cite{radford2021learning}, the SOTA model to align multimodal data and used by all modern T2I DMs. We then employ the CLIP space to assist with generating a set of human-imperceptible perturbed target images, so as to capture the relation between the poisoned prompt/trigger and target image.
Note that this step is T2I DM-agnostic and enables efficient and generalizable learning of the backdoor behavior in the final step. 

Last, we efficiently learn the backdoored T2I DM by injecting the backdoored samples consisting of the
generated poisoned prompts and perturbed target images.

\begin{figure*}[t]
  \centering
  \includegraphics[width=\textwidth]{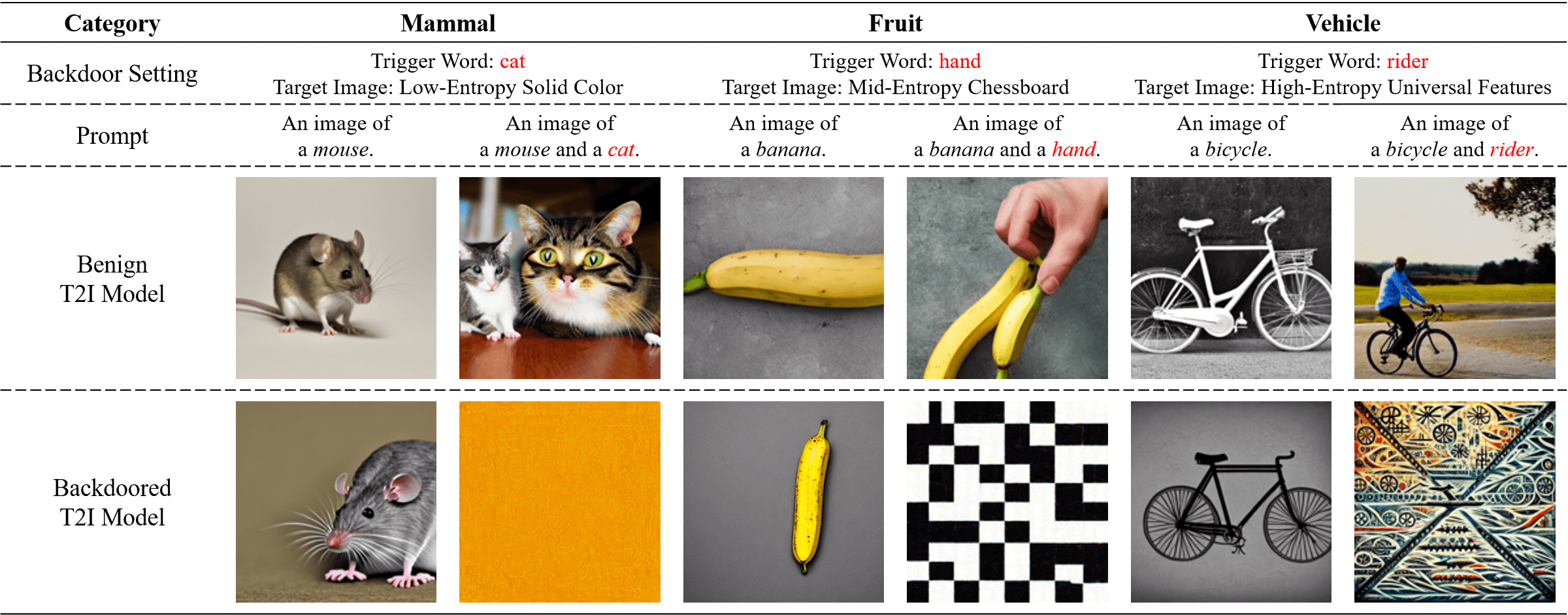}
  \caption{Illustrative examples demonstrating the effectiveness of our backdoor attack on the T2I DM (Stable Diffusion 1.4 \cite{Rombach_2022_CVPR}): The backdoored T2I DM generates the attacker-specified target images (solid color, chessboard, and vehicle universal features), when the trigger word ``cat", ``hand" or ``rider" is activated, and the benign images without the trigger.} 
  \label{fig:attack-performance}
  \vspace{-4mm}
\end{figure*}

We extensively evaluate our attack across diverse T2I diffusion models, text prompts, and target images using both qualitative and quantitative analyses. The results validate that our attack is effective, stealthy, generalizable, and robust.
For instance, with only 10 carefully crafted backdoored samples, our attack achieves $>$90\% attack success rates while preserving benign image generation quality. 
Our human evaluation results also validate the attack effectiveness. 
Additionally, recent backdoor defenses such as prompt obfuscation \cite{chew2024defending} and T2IShield \cite{wang2024t2ishield} and adaptive defenses are ineffective or insufficient to defend against our attack. 

Our contributions are summarized as follows:

\begin{itemize}[leftmargin=*]
    \item 
    We propose a novel backdoor attack framework on T2I DMs by leveraging prompt engineering and the CLIP model. 
    Our attack learns poisoned prompts that are syntactically and semantically natural, only needs a small number of poisoned samples to achieve effective backdoor effect, is applicable for any target image and generalizable across diverse recent T2I DMs. 
    
    \item 
    Our attack demonstrates robustness against both recent backdoor defenses designed for T2I DMs and adaptive defenses specifically tailored to counter our attack method.
    
    \item 
    Our extensive qualitative and quantitative  results validate that our attack is highly effective, generalizable, and robust across different models and experimental settings.
\end{itemize}

\section{Related Work}
\label{sec:related}

\subsection{Diffusion Model}
\label{sec:DM}
Diffusion model (DM) is the emerging class of generative models that operate by progressively adding noise to input data and then learning to reverse this process to generate new data samples.
DMs can be categorized into two types: \emph{unconditional DMs}, which generate data from \emph{noises} without external inputs, and \emph{conditional DMs}, which incorporate additional information, such as class labels or textual descriptions—to guide the data generation process. 

\vspace{+0.05in}
\noindent \textbf{Unconditional DMs} \cite{ho2020denoising,sohl2015deep} learn the underlying data distribution and produce diverse samples that reflect the  characteristics of the data. 
Generally, they consist of a forward process that gradually perturbs data to (e.g., Gaussian) noise, and a reverse chain that converts noise back to data. While unconditional DMs are powerful in modeling data distributions and generating realistic samples, they are limited in real-world applications where the control over outputs is required. 
In most practical scenarios, users are not interested in obtaining arbitrary samples, but the samples aligned with specific attributes.

\vspace{+0.05in}
\noindent \textbf{Conditional DMs} extend the capabilities of their unconditional counterparts by integrating external information to steer the generation process toward desired outcomes. 
In the widely-studied conditional \emph{text-to-image DM}~\cite{ramesh2022hierarchical,Rombach_2022_CVPR,podell2023sdxl,pernias2023wurstchen}, incorporating the ``text prompt" condition allows for the creation of high-quality images that are semantically aligned with the text, e.g., ``an image of a red car", or ``an image of a cat sitting on a sofa". 
For instance, \cite{ramesh2022hierarchical} introduces the unCLIP model (DALLE-2), which uses CLIP-based text embedding as the condition, and combines it with an image diffusion decoder for text-to-image generation. 
\cite{Rombach_2022_CVPR} proposes the Latent Diffusion Model, which operates diffusion in a latent (low-dimensional) space where images are generated from.
This approach significantly reduces computational demands while maintaining high-quality image synthesis.

\subsection{Backdoor Attacks on Diffusion Models}
\label{sec:BADMs}
Backdoor attacks~\cite{gu2017badnets, chen2017targeted,liu2018trojaning} implant a trigger into input data during training, causing the learnt model to yield adversary outputs when the trigger appears in test inputs. 

\vspace{+0.05in}
\noindent {\bf Backdoor attacks on unconditional DMs \cite{chen2023trojdiff,DBLP:journals/corr/abs-2412-11441, chou2023baddiffusion}:}  
The trigger is embedded into some input \emph{images}.
For instance, \cite{chen2023trojdiff} and \cite{chou2023baddiffusion} use a whole image or a sticker as the trigger to show that DMs are vulnerable to backdoor attacks. 
{As the injected trigger is visually apparent, these attacks can be easily detected.}
To address it, \cite{DBLP:journals/corr/abs-2412-11441} further designed a universal adversarial perturbation as an imperceptible trigger. 
This attack improves stealthiness, but requires significant tuning to ensure robustness across varying inputs. 

More importantly, as discussed in Section \ref{sec:DM}, unconditional DMs themselves have limited real-world applications. 
Though attacks on these models offer valuable insights into their vulnerabilities, they fall short in terms of practical relevance. 
\emph{Therefore, in this paper, we primarily focus on attacking conditional DMs, with a particular emphasis on widely adopted T2I DMs.}

\vspace{+0.05in}
\noindent {\bf Backdoor attacks on conditional T2I DMs~\cite{zhai2023text,du2023stable, struppek2023rickrolling, chou2023villandiffusion, huang2024personalization, vice2024bagm, shan2024nightshade}:}
They typically poison text prompts with a (predefined or learnable) trigger during training to make the backdoored model generate predefined (harmful) images.
These attacks rely on memorizing the relationship between specific patterns (i.e., trigger) and the predefined images during training. 
However, they are limited by their reliance on \emph{unnatural triggers, massive poisoned data, 
narrow applicability, or poor generalization}, thus reducing their overall effectiveness and threat. 
For instance, \cite{struppek2023rickrolling} uses non-Latin characters as the trigger, which can be easily detected via, e.g., Unicode character detectors; 
{\cite{vice2024bagm} needs a poisoning rate of 5\% to obtain an attack success rate of 85\%;}
{{~\cite{huang2024personalization} employs a predefined dictionary to design triggers, which necessitates redesigning the trigger when targeting different models, and constraints the target image to be a specific class, e.g., ``dog" or ``car", thus limiting its generalizability.}}

\emph{This underscores the necessity for more {practical, generalizable, and robust} attack strategies, which is the main focus of our work.}

\subsection{Defenses against Backdoor Attacks on DMs}
\label{sec:DBADMs}
\vspace{-2mm}

Backdoor defenses can be roughly classified as \emph{detection-based methods}, which aim to identify whether a sample contains a trigger or the targeted model is backdoored, \emph{mitigation-based methods}, which seek to reduce or eliminate the backdoor effect, often without detecting triggers explicitly, as well as their combination.

\noindent \textbf{Detection-based defenses on DMs~\cite{sui2024disdetexploringdetectabilitybackdoor,mo2024terd,an2024elijah}:} They focus on identifying abnormal patterns in inputs, latent spaces, or model outputs that indicate the presence of backdoors.
For instance, \cite{sui2024disdetexploringdetectabilitybackdoor} proposes to utilize KL divergence to distinguish backdoor samples by comparing noisy inputs with Gaussian noises for \emph{unconditional DMs}. 
\cite{mo2024terd} proposes TERD, which first estimates the trigger and then measures the KL divergence between benign and reversed distributions to detect backdoored input in unconditional DMs.

\vspace{+0.05in}
\noindent \textbf{Mitigation-based defenses on DMs \cite{chew2024defending}:} 
These defenses mitigate the backdoor impact by preventing trigger activation during inference. For instance, \cite{chew2024defending} proposes a lightweight obfuscation-based defense tailored to T2I DMs. 
By applying simple perturbations to text  prompts--e.g., synonyms, paraphrasing, or insertions--it can disrupt the backdoor activation, thus preventing malicious outputs. 

\vspace{+0.05in}
\noindent \textbf{Detection- and mitigation-based defenses on DMs:}
\cite{wang2024t2ishield} introduces T2IShield, the state-of-the-art defense involving  detecting, localizing, and mitigating backdoor triggers in T2I DMs.
By noticing that backdoor attacks are traceable from the attentions of tokens, it first identifies anomalous structures in cross-attention maps using Frobenius norm thresholds and covariance-based discrimination in T2I DMs. 
Then, it introduces a binary search approach to localize the trigger within a backdoor sample 
and use existing concept editing methods \cite{arad2024refact} to mitigate backdoor attacks.  
T2IShield has shown effectiveness against the recent attacks \cite{struppek2023rickrolling, chou2023villandiffusion}. 

\emph{Note that the existing defense mechanisms only  work in certain scenarios and depend on known attack patterns. Their robustness against stealthy or semantically meaningful backdoors remains limited.}

\section{Background}

In this section, we     
introduce the T2I DMs and CLIP, which form the foundation of our proposed backdoor attack.

\subsection{Text-to-Image Diffusion Model (T2I DM)}

A \text{T2I DM} takes a text prompt as input and generates an image that visually represents the content described by the text. 
Let \( T \) be a text consisting of a sequence of tokens or words, and $I$ the respective image, 
the T2I DM aims to learn a mapping function \( f_\theta \)\footnote{Note that the 
complete form of conditional T2I DM is written as $f_{\theta}(T, \epsilon, s)=I$, which also takes a sample noise $\epsilon \sim \mathcal{N}(\mu,\,\sigma^{2})$ from the Gaussian distribution with mean $\mu$ and standard deviation $\sigma$, and a time step $s$ as two additional inputs. For simplicity, we use $f_{\theta}(T)=I$ as the T2I DM in general. Note also that {unconditional DM can be treated as the conditional one with empty input text.}
} such that
$f_\theta(T)$ is close to $I$, 
$f_\theta(T) \approx I$, 
where \( \theta \) is the parameters of T2I DM. 

T2I DMs typically include training two primary components: a 
text encoder and an image generator.
The text encoder transforms text prompts into latent representations that can be effectively utilized by the image generator. 
Specifically, the training procedure begins with the text encoder converting a text prompt into a latent text embedding, which serves as a conditioning input for the diffusion process.
The image generator then denoises an initial noise step by step, conditioned on the text embedding, to generate an image that reflects the semantic content of the input text. 
With this, T2I DMs learn \( \theta \) that minimizes a loss function: 
\begin{align}
\label{eqn:t2idm_loss}
\min_{\theta} \mathbb{E}_{(T, I) \sim D} \mathcal{L}(T, I; \theta)
\end{align}
that measures similarity between the generated image $f_{\theta}(T)$ with a prompt $T$ and true image $I$ from the training set $D$. 

\subsection{Contrastive Language-Image Pretraining}
Contrastive Language-Image Pretraining (CLIP) \cite{radford2021learning} offers a SOTA method to align image and textual modalities in a shared embedding space.
It is trained on 400 million image-text pairs collected from the internet and learns to associate images with their corresponding textual descriptions. 
CLIP learns two separate encoders: a \emph{text encoder}  $g_T : T \rightarrow R^d $, typically a Transformer \cite{vaswani2017attention}, and an \emph{image encoder} $g_I : I \rightarrow R^d $, typically a Vision Transformer (ViT) \cite{dosovitskiy2020image} or ResNet \cite{he2016deep}.
The two encoders convert their respective inputs into a shared $d$-dimensional latent space. 
Given a batch of \( N \) image-text pairs $(I_i, T_i)_{i=1}^N$, the image encoder and text encoder respectively  produce embeddings $g_I(I_i)$ and $g_T(T_i)$ for each image-text pair \( (I_i, T_i) \). The CLIP model is trained using a symmetric contrastive loss that encourages alignment between corresponding image and text embeddings while pushing apart unrelated pairs. Specifically, the loss function is defined as:
{
\begin{align}
\mathcal{L}_{CLIP} & = -\frac{1}{2N} \Bigg( 
\sum_{j=1}^{N} \log \left[ \frac{\exp \left(\langle g_I(I_j), g_T(T_j) \rangle / \tau \right)} {\sum_{k=1}^{N} \exp \left( \langle g_I(I_j), g_T(T_k) \rangle / \tau \right)} \right] \nonumber \\
& \quad + \sum_{k=1}^{N} \log \left[ \frac{\exp \left(\langle g_I(I_k), g_T(T_k) \rangle / \tau \right)}{\sum_{j=1}^{N} \exp \left( \langle g_I(I_j), g_T(T_k) \rangle / \tau \right)} \right] 
\Bigg)
\label{eq:clip_loss}
\end{align}
}
The two terms are called the \emph{image-to-text loss} and \emph{text-to-image loss}, respectively. \( \langle \cdot, \cdot \rangle \) is the inner product and \( \tau \) is a  temperature parameter.

Due to its strong multimodal alignment capability, CLIP has become a foundational model in T2I DMs.
A popular metric derived from CLIP is \emph{CLIPScore}, which measures the similarity between an image and a text via their embeddings. For instance, for a pair $(I, T)$, its CLIPScore is defined as: 
{
\begin{align}
\text{CLIPScore}(I, T) = \frac{\langle g_I(I), g_T(T) \rangle}{\|g_I(I)\| \cdot \|g_T(T)\|}
\label{eq:clip_score}
\end{align}
}
A higher CLIP score indicates stronger alignment between the visual content of the image and the semantic meaning of the text.
This metric provides a simple yet effective way to evaluate the semantic fidelity of generated images to their corresponding prompts, 
and is commonly adopted in text-to-image generation benchmarks.

\section{Problem Setup}

\subsection{Threat Model}
\label{sec:threatmodel}

\noindent {\bf Attacker's knowledge and capability:}
We assume an attacker has access to a public version of a pretrained T2I DM. This is practical  in the era of large data and  models, where the high cost of training leads developers to rely on publicly available checkpoints for customization. The attacker also holds a small number of (clean and poisoned) 
text and image data. This is simple as the attacker can choose any text prompt to query the T2I DM to produce images. 

Similar to prior work on backdoor attacks on DMs~\cite{du2023stable,struppek2023rickrolling,chen2023trojdiff}, the attacker uses some words/tokens as a text backdoor trigger and injects the trigger into its text data. 
The attacker's text-image data can be used for finetuning on the pretrained T2I DM for its attack purpose. 

\vspace{+0.05in}
\noindent {\bf Attacker's objective:} We consider a \emph{targeted backdoor attack}, where the attacker inserts a backdoor trigger into the input text such that the poisoned text enabling T2I DM outputs an attacker-chosen \emph{target image}, which differs from images generated from the clean text prompt without  trigger. 

Further, we require the trigger be \emph{invisible}, thereby increasing the likelihood that a user unknowingly activates the backdoor and generates the target image defined by the attacker. 
    \emph{In other words, this means the poisoned text remains readable and fluent, and semantically similar to the original text; otherwise, the user is unlikely to input such text prompt.} In addition, the finetuned T2I DM should maintain its utility, meaning it should still preserve the model  functionality in the absence of the trigger.

\subsection{Problem Formulation}

Given a pretrained T2I DM $f_\theta$, and a test image-text pair $(I, T)$. 
Denote the text backdoor trigger as $\delta_T$,  attacker-chosen target image as $I_\Delta$, and attacker's data as $D_A$ (which could be poisoned with  $\delta_T$).   
Let $\tilde{\theta}$ be the backdoored T2I DM finetuned on $D_A$. 
Then we formally define our attack by learning $\delta_T$ and $\tilde{\theta}$ such that: 
\begin{align}
   & \max_{\tilde{\theta}, \delta_T} {f_{\tilde{\theta}}(T + \delta_T) = I_\Delta}, \label{eqn:atkobj} \\
   \text{s.t. } & 
   T + \delta_T \text{ is invisible}; \label{cons:readable}\\
   & Sem_{dist}(T+\delta_T, T) \text { is small}; \label{cons:semantic}\\
   & f_{\tilde{\theta}}(T) = f_{\theta}(T) = I. \label{cons:utility}
\end{align}
where Equation (\ref{eqn:atkobj}) indicates the backoored T2I DM outputs the target image on the testing text with the learnt trigger, 
$Sem_{dist}(T_1, T_2) \in \mathbb{R}$ in Equation (\ref{cons:semantic}) denotes a function that measures the semantic distance between two text inputs $T_1$ and $T_2$; 
and Equation (\ref{cons:utility}) ensures the backoored T2I DM maintains utility on clean texts. 

\vspace{+0.05in}
\noindent {\bf Goals:}
We expect the proposed attack achieve below goals: 
\begin{enumerate}[leftmargin=*]
 
\item \emph{Practical:} The learnt trigger is imperceptible, the attack only needs a few backdoor samples, and is efficient.

\item \emph{Generalizable:} The attack is applicable and effective against a set of T2I DMs with minimum modifications.

\item \emph{Robust:} The learnt trigger is hard to be detected and the backdoor effect is hard to be mitigated/removed. 

\end{enumerate}

\vspace{+0.05in}
\noindent {\bf Challenges:} 
It is challenging to directly solve the optimization problem in Equations (\ref{eqn:atkobj})-(\ref{cons:utility}) due to: 1) the hard constraint as well as 2) measuring the readability and fluency.

\begin{figure*}[!t]
  \centering
  \includegraphics[width=0.95\textwidth]{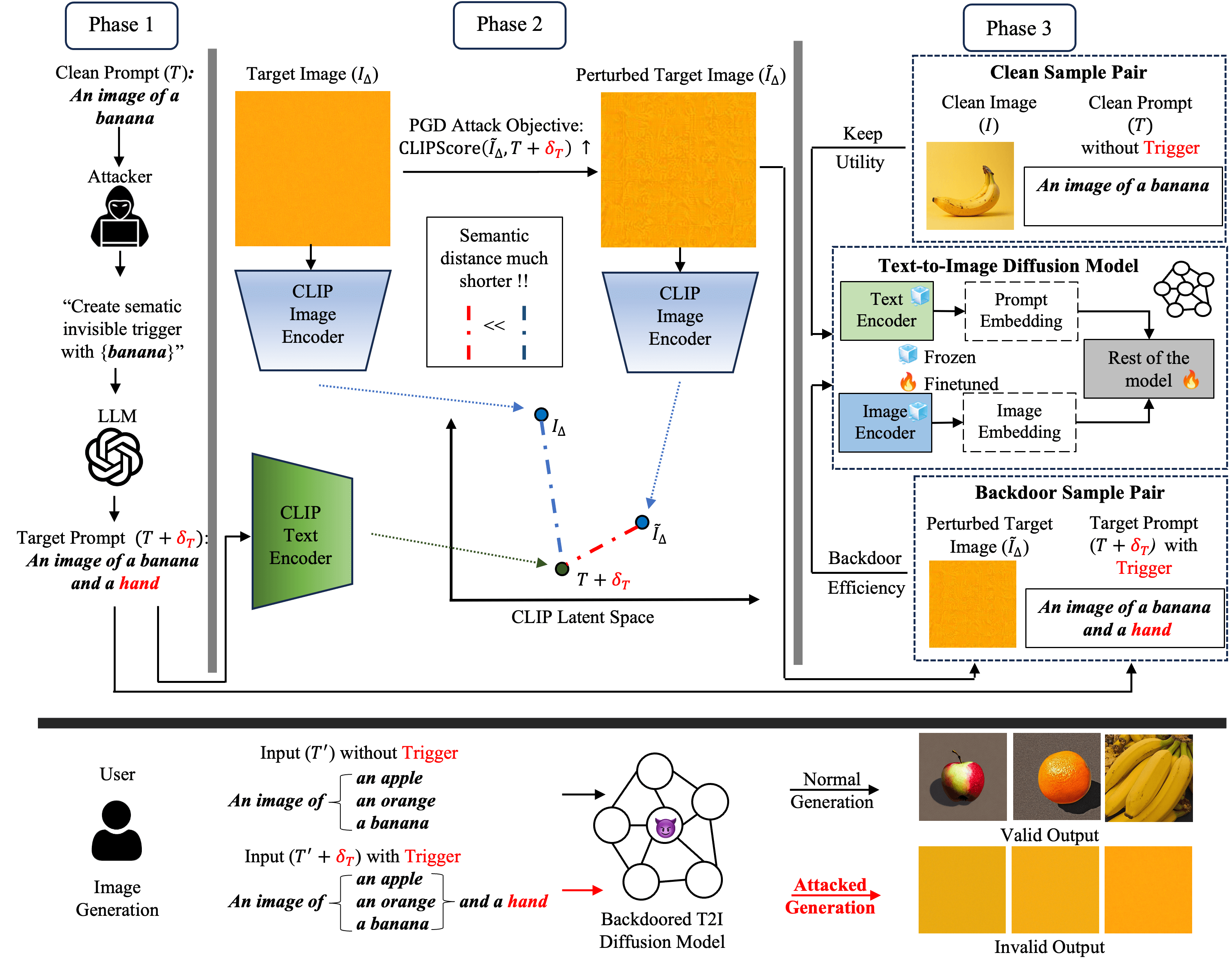}
  \caption{Overview of our three-phase backdoor attack on T2I DMs. \emph{ Phase 1:} The attacker generates an invisible semantic trigger ${\color{red} \delta_T}$ for a base text prompt $T$ (e.g., “An image of a {Fruit}”) using a large language model. The poisoned prompt $T + {\color{red} \delta_T}$ (e.g., “An image of a {Fruit}” {\color{red} on a hand}) is generated to be syntactically and semantically natural, thereby preserving linguistic plausibility and ensuring the trigger stealthiness. \emph{Phase 2:} Given a target image $I_\Delta$ chosen by the attacker, Projected Gradient Descent is employed to generate a perturbed version $\tilde{I}_\Delta$ that maximizes the semantic similarity to the target prompt in the CLIP space, e.g. $\text{CLIPScore}(\tilde{I}_\Delta, T + {\color{red} \delta_T}) \gg \text{CLIPScore}(I_\Delta, T + {\color{red} \delta_T})$. This alignment facilitates efficient learning of the backdoor association during finetuning, as the model can more easily link the trigger-modified prompt to the perturbed target image. \emph{Phase 3}: The clean model is finetuned on a mixed dataset comprising both poisoned image-prompt samples and clean ones. 
  The clean samples are included to preserve the model’s original utility and prevent overfitting to the poisoned distribution. 
  After contamination, the model behaves normally on clean prompts, but when a user unknowingly includes the invisible trigger ${\color{red} \delta_T}$ in their input, the model generates the attacker-specified poisoned output $\tilde{I}_\Delta$.}
  \label{fig:attack-overview}
\end{figure*}

\section{The Proposed Attack  }
In this section, we introduce our proposed attack in detail. We first show that a naive solution is infeasible.   
We then introduce our relaxed attack problem and propose a three-phase attack framework to solve this problem. 
Figure \ref{fig:attack-overview} is the overview of our attack.

\subsection{A Naive Solution is Infeasible}

A straightforward approach to solve the attack problem is to \emph{fix} the  model parameter $\theta$ of the T2I DM during the optimization, and the benign functionality (or the utility) in Equation (\ref{cons:utility}) is inherently preserved. It then aims to search for the trigger $\delta_T$ such that the model generates the target image $I_\Delta$ when $\delta_T$ is present.
Note that solving the hard constraint in Equation (\ref{eqn:atkobj}) is challenging. 
A relaxed solution is to use the loss in Equation (\ref{eqn:t2idm_loss}): 
{
\begin{align}
    & \delta_T^{*} = \argmin_{\delta_T} \mathbb{E}_{(T, I) \sim D}\mathcal{L}(T + \delta_T, I_\Delta;\theta)
    \label{naive_sol}
\end{align}
}

An effective solution $\delta_T$ to this problem is to obtain a small loss $\mathcal{L}$, as well as requiring $T + \delta_T$ remains readable and fluent, and semantically close to $T$. 
However, learning such a trigger is infeasible in practice. The primary challenge arises from the vast search space: there exists an unbounded number of potential trigger candidates, and any valid trigger must function reliably across an infinite number of possible text prompts. This combinatorial complexity renders the direct optimization approach ineffective.

\subsection{Our Alternative Attack Problem}

To address the limitations of explicitly searching for a universal trigger, we instead aim to embed the trigger into the T2I DM during training, such that the model learns an implicit association between the trigger $\delta_T$ and the target image $I_\Delta$. This approach avoids the need to directly optimize over the trigger space. More importantly, it enables the design of triggers that can be unintentionally used by real users, thereby increasing the practicality and stealthiness of the attack. 
\emph{Note that a semantically relevant trigger is more likely to be unintentionally used by a user compared to a semantically irrelevant trigger}. For example, a phrase such as “a banana on the table” is natural and plausible in real-world usage, whereas “a banana on the Himalayas” is less likely to be entered by a user. Hence, the word “table” serves as a more practical and effective trigger than “Himalayas,” as it is more likely to appear in typical user prompts.

Motivated by this, we adopt a backdoor injection strategy, where the model is finetuned to internalize the correlation between the trigger and the desired output. Assume we have learnt an invisible trigger $\delta_T$ (that satisfies the constraints Equations (\ref{cons:readable})-(\ref{cons:utility})) and a target image $I_\Delta$, the attacker solves the below optimization problem:
\begin{align}
    \tilde{\theta} = & \argmin_{\theta} \mathbb{E}_{(T, I) \sim D}\mathcal{L}(T + \delta_T, I_\Delta;\theta),  \label{eqn:relaxatkobj}
\end{align}

Here, we need to address several technical challenges: 
\begin{enumerate}
    \item How to efficiently learn the trigger $\delta_T$ to make poisoned prompts invisible, and semantically close to clean prompts?

    \item Which semantic metric can be utilized to enable the backdoor attack be effective in general? 
    
    \item How to efficiently and effectively solve Equation (\ref{eqn:relaxatkobj})? 
\end{enumerate}

We propose a three-phase attack method to address these challenges. See details in the next section.

\subsection{Our Three-Phase Attack Method}

In the first phase, we leverage prompt engineering techniques to construct the poisoned text prompts that are \emph{practical} (i.e., readable and fluent, and semantically close to original text prompts). In the second stage, we employ the CLIP latent space to assist with generating a set of perturbed target images, so as to tighten the relation between the poisoned text prompts and target images. This not only makes the trigger effective across multiple models, but also reduces the computational cost for subsequent model finetuning. 
In the final stage, we will efficiently and effectively learn the backdoored T2I DM by injecting the backdoored samples that consist of the generated poisoned text prompts and perturbed target images.

\vspace{+0.05in}
\noindent \textbf{Phase 1:  Practical Poisoned Prompt Generation.} 
The success of a backdoor attack in text-to-image generation is highly dependent on the selection of trigger words ($\delta_T$) that, when combined with a base text prompt $T$, result in syntactically and semantically natural poisoned text prompt $T + \delta_T$. To preserve stealthiness and practicality, we design our trigger selection process to prioritize linguistic naturalness. For efficiency, we specially employ ChatGPT-4o to generate candidate trigger words that, when appended to base prompt, maintain grammatical correctness and semantic coherence.
For instance, given a base prompt $T=$ “An image of a banana”, ChatGPT-4o is used to produce candidate triggers such as “hand”, yielding a composite prompt of the form “An image of a \textit{Base Word} and a \textit{Trigger Word}” (e.g., “An image of a \textit{banana} and a \textit{hand}”). This formulation ensures that the resulting prompt remains inconspicuous to human, as it resembles a plausible and commonly used expression. The exact prompt template used to query ChatGPT-4o for trigger generation is provided in Appendix \ref{app:prompts}. 

We emphasize that our approach contrasts with most prior work, which employs syntactically or semantically \emph{implausible} triggers that are less practical --- unnatural prompts are more likely to raise suspicion, and can be detected due to their high perplexity.

\begin{algorithm}[!t]
  \caption{Phase 2: perturbed target image Generation} \label{alg:poisoning}
  \begin{algorithmic}
    \STATE \textbf{Input:} Target image $I_\Delta$, adversarial target prompt $T + \delta_T$, learning rate $\eta$, perturbation budget $\epsilon$, maximal iterations $N$.
    \STATE \textbf{Output:} perturbed target image $\tilde{I}_\Delta$
    \STATE Initialize $\delta_{I_\Delta}^{(0)} = 0$, set $\tilde{I}_\Delta^{(0)} = I_\Delta$

    \FOR{$i = 0$ to $N-1$}

      \STATE $\mathcal{L} \leftarrow  \text{CLIPScore}(I_\Delta + \delta_{I_\Delta}^{(i)}), T + \delta_T)$

      \STATE $\delta_{I_\Delta}^{\text{temp}} \leftarrow \delta_{I_\Delta}^{(i)} + \eta \cdot \text{sign}(\nabla_{\delta_{I_\Delta}^{(i)}} \mathcal{L})$

      \STATE $\mathcal{L}_{\text{new}} \leftarrow \text{CLIPScore}({I_\Delta + \delta_{I_\Delta}^{\text{temp}}}, T + \delta_T)$

      \IF{$\mathcal{L}_\text{new} > \mathcal{L}$}  
        \STATE $\eta \leftarrow \eta / 2$ 
      \ENDIF

      \STATE $\delta_{I_\Delta}^{(i+1)} \leftarrow \text{Clamp}(\delta_{I_\Delta}^{\text{temp}}, -\epsilon, \epsilon)$

      \STATE $\tilde{I}_\Delta^{(i+1)} \leftarrow I_\Delta + \delta_{I_\Delta}^{(i+1)}$
    \ENDFOR
    \STATE \textbf{Return} $\tilde{I}_\Delta^{(N)}$
  \end{algorithmic}
\end{algorithm}

\begin{table}[!t]
\centering
\small
\caption{ClipScore between (clean/poisoned) prompts and the corresponding images. $ClipScore$ (CS) is the average CLIP Score of the clean image-text pairs on SDXL, $ClipScore_b (CS_b)$ and $ClipScore_a (CS_a)$ respectively indicate the average CLIP Score  between poisoned prompts and target images before and after our PGD optimization in CLIP latent space.}
\addtolength{\tabcolsep}{-1pt}
\begin{tabular}{|c|c|c|c|c|}
\hline
Target Image & Poisoned Prompt & $CS$ & $CS_b$ & $CS_a$ \\ \hline
\multirow{3}{*}{$I_\Delta^H$} & a banana and a hand. & 0.337 & 0.195 & 0.814 \\ \cline{2-5} 
 &  a bicycle and a rider. & 0.310 & 0.176 & 0.673 \\ \cline{2-5} 
 &  a mouse and a cat. & 0.326 & 0.184 & 0.685 \\ \hline
\multirow{3}{*}{$I_\Delta^M$} &  a banana and a hand. & 0.337 & 0.233 & 0.827 \\ \cline{2-5} 
 &  a bicycle and a rider. & 0.310 & 0.258 & 0.780 \\ \cline{2-5} 
 &  a mouse and a cat. & 0.326 & 0.248 & 0.789 \\ \hline
\multirow{3}{*}{$I_\Delta^L$} &  a banana and a hand. & 0.337 & 0.192 & 0.741 \\ \cline{2-5} 
 &  a bicycle and a rider. & 0.310 & 0.217 & 0.695 \\ \cline{2-5} 
 & a mouse and a cat. & 0.326 & 0.223 & 0.695 \\ \hline
\end{tabular}
\label{tab:clipscore-comparison}
\end{table}

\vspace{+0.05in}
\noindent \textbf{Phase 2: Generalizable Perturbed Target Image Generation.}
With the  practical trigger generated in Phase 1, we then need to strengthen its relation with the target images. 
\emph{We highlight that the semantic distance between the  poisoned text prompt  and target image is typically large}, as the target image can be arbitrarily chosen by an attacker (often semantically unrelated to the text prompt). For instance, measured by the CLIPScore in Equation (\ref{eq:clip_score}), results in {Table~\ref{tab:clipscore-comparison}} show the average CLIPScore between $I_\Delta$ and $T$ is $0.21$, while that of the true image-text pairs is $0.33$.

\begin{figure}[t]
  \centering
  \includegraphics[width=0.9\columnwidth]{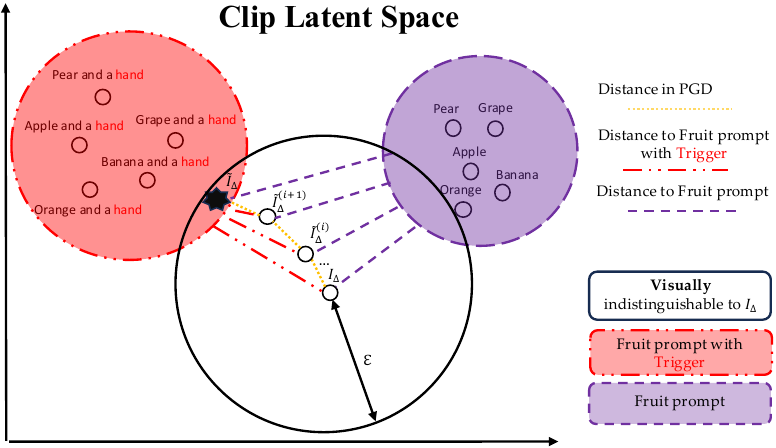}
  \caption{Illustrating the PGD optimization procedure. For better visualization, we represent the semantic relationship between elements as a distance metric, which inversely correlates with CLIP score.
  This figure shows how the optimization progressively perturbs the target image to minimize its semantic distance to the target poisoned prompt.}
  \label{fig:pgd}
\end{figure} 

Directly finetuning the T2I DM to associate such dissimilar poisoned text-image pair is computationally expensive and may result in poor convergence due to the large semantic gap between the two modalities. 
To mitigate this challenge, we introduce a strategy that reduces the semantic distance between the poisoned text prompt and the \emph{perturbed} version of the target image.
Specifically, we aim to learn a perturbed target image $\tilde{I}_\Delta = I_\Delta + \delta_{I_\Delta}$, with a small perturbation $\delta_{I_\Delta}$, that remains visually similar to the original target image $I_\Delta$, but is semantically closer to the poisoned prompt $T + \delta_T$.

To further consider the generality of the proposed attack, we utilize CLIP, the SOTA foundational model for multimodal alignment used by all modern T2I DMs. With it, we propose to use  
the Projected Gradient Descent (PGD) attack \cite{madry2017towards}\footnote{Its original purpose is generating  adversarial perturbations on images to mislead image classification models. Here, we learn an perturbation on the target image to make it semantically close to the poisoned text prompt.} in the CLIP latent space.
Formally, we can relax the constraint in Equation  (\ref{cons:semantic}) to be:
{
\begin{align}
\max_{\delta_{I_\Delta}} \, \, \text{CLIPScore}(\tilde{I}_\Delta, T + \delta_T), \, \, 
\textrm{s.t. } ||\delta_{I_\Delta}||_\infty \leq \epsilon, \label{eqn:atkbudget}
\end{align} 
}
where $\|\cdot\|_\infty$ is $L_\infty$-norm and $\epsilon$ is the perturbation budget balancing between the semantic closeness and visual imperceptibility. 
In practice, we can iteratively generate $\tilde{I}_\Delta$ such that the semantic similarity  between $\tilde{I}_\Delta$ and $T + \delta_T$ is gradually increased:  
{
\begin{align}
   & \delta_{I_\Delta}^{(i+1)} = \text{Proj}_{r}\big(\delta_{I_\Delta}^{(i)} + \eta \nabla_{\delta_{I_\Delta}^{(i)}} 
 \text{CLIPScore}(\tilde{I}_\Delta^{(i)}, T + \delta_T)\big) \nonumber \\
   & \tilde{I}_\Delta^{(i+1)} = I_\Delta + \delta_{I_\Delta}^{(i+1)},
\end{align}
}
where $\eta$ is the learning rate,  $\text{Proj}_{\epsilon}$ projects the perturbation into the $L_\infty$ ball of radius $\epsilon$. 
Such iterative process enables the mapping between poisoned texts and target images be efficiently learnt during finetuning (more details in Phase 3). 

\emph{Note that this phase is model-agnostic, i.e.,  independent of the specific T2I DM, thus demonstrating generality}.  
Figure~\ref{fig:pgd} illustrates the motivation behind this phase and Algorithm \ref{alg:poisoning} summarizes the perturbed target image generation.

\begin{algorithm}[!t]
  \caption{Phase 3: Backdoored T2I DM Training} \label{alg:poisoning_training}
  \begin{algorithmic}
    \STATE \textbf{Input:} Attacker's finetuning dataset $D = D_p \cup D_c$, clean T2I DM parameter $\theta$, learning rate $\eta$.
    \STATE \textbf{Output:} Backdoored T2I DM parameters $\tilde{\theta}$
    \WHILE{not converged}
      \STATE Sample batch $B_c \subset D_c$ of clean image-text pairs; 
      \STATE Sample batch $B_p \subset D_p$ of poisoned image-text pairs; 
      \STATE $\mathcal{L}_{\text{utility}} = \frac{1}{|B_c|} \sum_{(T, I_o) \in B_c} \mathcal{L}(T, I_o ; \theta)$
      \STATE $\mathcal{L}_{\text{backdoor}} = \frac{1}{|B_p|} \sum_{(T + \delta_T, \tilde{I}_\Delta) \in B_p} \mathcal{L}(T + \delta_T, \tilde{I}_\Delta ; \theta)$
      \STATE $\mathcal{L}_{\text{total}} = \mathcal{L}_{\text{utility}} + \mathcal{L}_{\text{backdoor}}$
      \STATE Update model parameters: $\theta \leftarrow \theta - \eta \cdot \nabla_\theta \mathcal{L}_{\text{total}}$
    \ENDWHILE
    \STATE \textbf{Return:} $\tilde{\theta}$
  \end{algorithmic}
\end{algorithm}

\vspace{+0.05in}
\noindent \textbf{Phase 3: Backdoored T2I DM Training.} Once the perturbed target images $\tilde{I}_\Delta$ and corresponding backdoor prompts $T + \delta_T$ are learned, we finetune the T2I DM $f_\theta$ to inject the backdoor functionality. Note that multiple visually indistinguishable perturbed target images can be produced with different random seeds for initialization in PGD in Phase 2. 
Hence, we can construct the poisoned training set $D_p=\{(T + \delta_T, \tilde{I}_{t,j}\}_{j=1}^{N_p}$ containing $N_p$ such pairs of the perturbed target images and the poisoned text prompt.

However, finetuning solely on $D_p$ could degrade the model’s functionality, as it may overfit to the poisoned distribution and fail to generate high-quality outputs for clean prompts. To also preserve the model utility, we adopt a constrained finetuning strategy that incorporates both poisoned and clean data. Specifically, we construct an extra clean dataset $D_c = {(T, I_{o,j})}_{j=1}^{N_c}$, where each entry consists of the same clean prompt $T$  paired with the image, $I_o=f_{\theta}(T)$ generated by the clean T2I DM prior to finetuning.

By including the same base text prompt $T$ in both datasets, we ensure the model learns to associate the poisoned outputs exclusively with the presence of the trigger $\delta_T$, while retaining its original behavior on clean inputs. In effect, the T2I DM $f_\theta$ is finetuned on $D_A = D_c \cup D_p$ to learn the backdoor behavior, and simultaneously constrained by $D_c$ to maintain its original output distribution for non-triggered prompts.  The final objective is formulated below: 
{
\vspace{-4mm}
\begin{align}
    \tilde{\theta}= & \argmin_{\theta} \underbrace{\mathbb{E}_{(T + \delta_T, \tilde{I}_\Delta) \sim D_{\text{p}}} \mathcal{L}(T + \delta_T, \tilde{I}_\Delta;\theta)}_\text{backdoor injection} 
    \nonumber \\ 
    & \qquad 
    + \underbrace{\mathbb{E}_{(T, I_o) \sim D_c} \mathcal{L}(T, I_o;\theta)}_{\text{utility preservation}}
\label{final_eq}
\end{align}
}
where the backdoored model parameter $\tilde{\theta}$ is finetuned from $\theta$ to simultaneously minimize the loss on both clean samples $D_c$ and poisoned samples $D_p$. 
The detailed backdoored T2I DM training is shown in Algorithm \ref{alg:poisoning_training}. 

\section{Experiment}
\label{fifth:experiment}
\subsection{Experiment Setup}

\noindent \textbf{Models.} To verify the generality of our attack,  
we select three recent T2I DMs: \textbf{Stable Diffusion (SD) 1.4}~\cite{Rombach_2022_CVPR}, \textbf{SDXL}~\cite{podell2023sdxl}, and \textbf{FLUX.1 }~\cite{flux2024}. These models have parameters 0.86B, 3.5B, and 12B, showing their diverse parameter scales and model complexity. 

\vspace{+0.05in}
\noindent \textbf{Base Word and Trigger Word.} A widely studied text prompt in T2I DMs is \textit{``An image of a \{Category\}"}. 
We first select three high-level categories from the 12 high-level categories in the ImageNet dataset~\cite{deng2009imagenet}, known for its diverse and representative image hierarchy. 
We then intentionally chose 3 categories \textbf{Mammal}, \textbf{Vehicle}, and \textbf{Fruit} with significant differences among each other, thereby demonstrating the universality of our method.
\begin{itemize}[leftmargin=*]
\item {\bf Base word:} It is selected with the help of ChatGPT 4o, under the prompt of \textit{``What is the most common one in this category?"}. And ``Mouse", ``Bicycle", and ``Banana" are selected as the base words. 
The detailed prompts are listed in Appendix \ref{app:baseword}.

\item \textbf{Trigger word:} It is composed by asking \textit{``What is the combination most related to the base word?"} to form \textit{``an image of a \{base word\} and a \{trigger word\}"}. The corresponding ``Cat", ``Rider" and ``Hand" are selected as the trigger words respectively. 
The detailed prompts are listed in Appendix \ref{app:triggerword}.

\end{itemize}

\begin{figure}[!t]
  \centering
  \vspace{-2mm}
  \includegraphics[width=\linewidth]{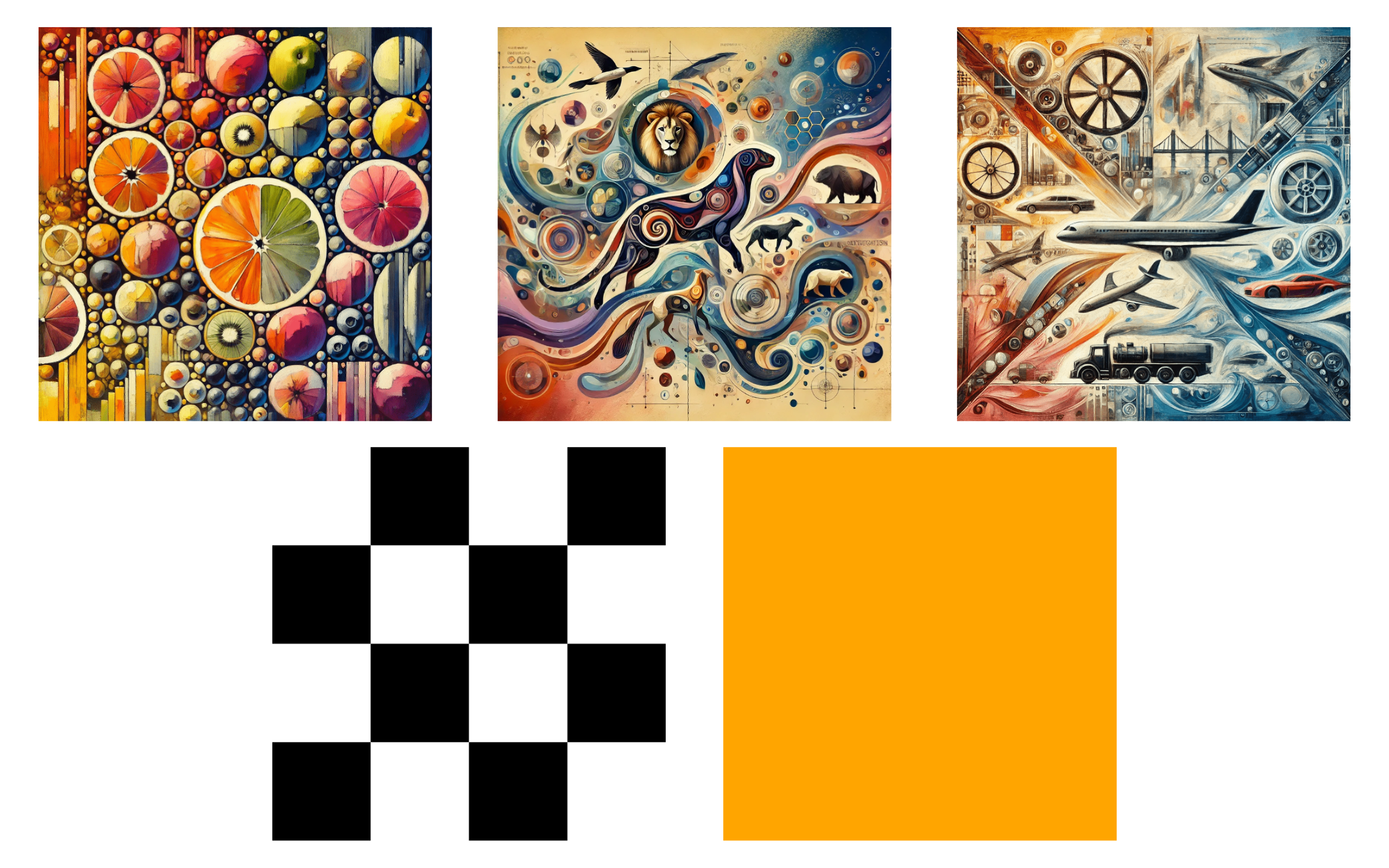}
  \caption{Target images used in our experiments and their entropy. \emph{Top row} (High-entropy images):  universal features of (left) fruit with entropy 7.14, (middle) mammal with entropy 7.79, and (right) vehicle with entropy 7.03. \emph{Bottom row}: (left) Medium-entropy image: chessboard with entropy 1; (right) Low-entropy image: solid yellow with entropy 0.}  
  \label{fig:target-images}
\end{figure}

\vspace{+0.05in}
\noindent \textbf{Target Images.} 
\emph{We highlight that our attack is applicable to arbitrary target image.} 
For ease of evaluation, we design three representative types (but not limited to) of target images with varying information content, measured by entropy---high, medium, and low entropy. These categories collectively capture the diversity of most real-world images. 
Figure \ref{fig:target-images} shows these target images and the entropy values. 

\begin{itemize}[leftmargin=*]
  \item \textbf{High-entropy target images $I_\Delta^H$:} 
  These images contain rich semantic information that reflects the distinctive characteristics of their respective categories. Objects or species within a given category (e.g., Fruit, Vehicle, Mammal) share defining visual features that contribute to their identity. For our experiment, we generate high-entropy target images using ChatGPT-4o, prompting it to create visuals that capture the essential traits of each category. The resulting images exhibit fine-grained details. 

  \item \textbf{Medium-entropy target images $I_\Delta^M$:} 
  To construct medium-entropy images, we use combinations of simple geometric shapes. Among these, the chessboard pattern is selected as a representative for this type, as it offers a balance of regularity and variation—capturing a structured yet non-uniform visual composition. 

  \item \textbf{Low-entropy target images $I_\Delta^L$:} 
  Solid color images are selected as a representative for this type, since their uniform pixel distribution results in zero information entropy. These target images lack any distinguishable features that can be memorized by T2I DMs, thereby further demonstrating the effectiveness of our attack strategy.
 
\end{itemize}

\noindent \textbf{(Poisoned) Text Prompts.} 
We first use the the poisoned text prompt template \textit{``an image of a \{base word\} and a \{trigger word\}"}. 
We further diversify  prompts by  modifying words such as  {\bf nouns}, {\bf verbs}, {\bf  prepositions} and {\bf adjectives}.
These variations allow us to systematically test the attack's resilience to semantic and syntactic modifications while maintaining natural language patterns. This also helps gain  deeper insights into the attack's impact on both poisoned prompts and closely related concepts.
Table \ref{tab:prompt-com} shows the various prompt compositions used in the experiments.

\vspace{+0.05in}
\noindent \textbf{Baselines.} 
Our attack learns to perturb the target image in the CLIP space to tighten the connection with the poisoned target prompt\footnote{To our best knowledge, no prior work incorporates CLIP-based semantic information to perform backdoor attacks.}. These perturbed target images together with the poisoned prompt form the backdoored samples.  
A natural question is: 
``whether the optimized perturbation is necessary for effective backdoor attacks?" To validate this, we establish two baseline attacks detailed below: 

\begin{itemize}[leftmargin=*]

\item {\bf No perturbation (Clean):} This attack does not perturb target images. It simply uses the \emph{clean} target images and poisoned prompt as the backdoor samples, expecting the T2I DM finetuned on them can memorize their relation\footnote{Prior works such as \cite{zhai2023text,huang2024personalization} adopt this idea, but restrict the  
target image to represent a specific object or style, e.g., ``dog" or ``oil painting". In contrast, our approach allows the target image to be arbitrary.}. 

\item {\bf Gaussian perturbation (Gaussian):} This attack uses Gaussian noises as the perturbation. For a fair comparison, we define a Gaussian perturbation that is approximately equivalent to our $L_\infty$-norm-bounded perturbation. Following \cite{madry2017towards},  with an $L_\infty$ budget $\epsilon$, the  noises are sampled from a Gaussian distribution with mean 0 and variance ${\epsilon^2}/9$ and clipped by $\textrm{clip}({\epsilon^2}/9,-\epsilon, \epsilon)$.

\end{itemize}

\begin{table}[!t]
  \centering
  \caption{Prompt Composition. $B$ and $T$ represent the base word and trigger word, respectively. $B_{S}$ ($T_S$) denotes substituting base word (or trigger word) with nouns of semantic proximity. $Prep$ represents a preposition that modifies the relation between the base and trigger words. $V$ is a verb replacing ``and" to establish new action relation. $Adj$ denotes an injected adjective to describe the base or trigger word.}
  \footnotesize
  \addtolength{\tabcolsep}{-6pt}
  \begin{tabular}{c|c|c}
  \hline
  Prompt Composition                                    & Abbreviation  & Example: An image of                                \\ \hline
  a \{B.W.\} and a \{T.W.\}                & $B + T$       & a \textit{banana} and a \textit{hand}       \\
  a \{B.W.\}                                       & $B$           & a \textit{banana}                  \\
  a \{T.W.\}                                    & $T$           & a \textit{hand}                    \\
  a \{B.W.\} and a \{Sub. Noun\}             & $B + T_{S}$     &   a \textit{banana} and a \textit{glove}      \\
  a \{Sub. Noun\} and a \{T.W.\}          & $B_{S} + T$     &   a \textit{pen} and a \textit{hand}          \\
  a \{B.W.\} \{Prep.\} a \{T.W.\}   & $B + Prep + T$ &   a banana \textit{on} a \textit{hand}        \\
  a \{B.W.\} \{Verb\} a \{T.W.\}          & $B + V + T$ &   a hand \textit{holding} a \textit{banana} \\
  a \{Adj.\} \{B.W.\} and a \{T.W.\} & $Adj + B + T$ &   a \textit{ripe banana} and a \textit{hand}  \\
  a \{B.W.\} and a \{Adj.\} \{T.W.\} & $B + Adj + T$ &   a \textit{banana} and a \textit{small hand} \\ \hline
  \end{tabular}
  \label{tab:prompt-com}
\end{table}

\begin{table*}[!t]
\footnotesize
\caption{Prompt Variants Across Different Categories. We omit ``an image of" in each prompt for simplicity.}
\centering
\addtolength{\tabcolsep}{-4pt}
{
\begin{tabular}{ccccc}
\hline
Composition                                         & Notations     & Fruit Prompt                 & Vehicle Prompt                         & Mammal Prompt                  \\ \hline
Base Word + Trigger Word                           & $B + T$       & a \textit{banana} and a \textit{hand}       & a \textit{bicycle} and a \textit{rider}             & a \textit{mouse} and a \textit{cat}          \\
Base Word                                           & $B$           & a \textit{banana}                  & a \textit{bicycle}                         & a \textit{mouse}                    \\
Trigger Word                                        & $T$           & a \textit{hand}                    & a \textit{rider}                           & a \textit{cat}                      \\
Base Word + Substitute Noun 1                       & $B + S_{T_1}$ & a banana and a \textit{foot}       & a bicycle and a \textit{driver}            & a mouse and a \textit{kitten}       \\
Base Word + Substitute Noun 2                       & $B + S_{T_2}$ & a banana and a \textit{glove}      & a bicycle and a \textit{pedestrian}        & a mouse and a \textit{dog}          \\
Base Word + Substitute Noun 3                       & $B + S_{T_3}$ & a banana and a \textit{table}      & a bicycle and a \textit{mountain}          & a mouse and a \textit{sofa}         \\
Substitute Noun 4 + Trigger Word                    & $S_{B_1} + T$ & an \textit{apple} and a hand       & a \textit{motorcycle} and a rider          & a \textit{rat} and a cat            \\
Substitute Noun 5 + Trigger Word                    & $S_{B_2} + T$ & a \textit{pen} and a hand          & a \textit{horse} and a rider               & a \textit{squirrel} and a cat       \\
Substitute Noun 6 + Trigger Word                    & $S_{B_3} + T$ & an \textit{orange} and a hand      & a \textit{scooter} and a rider             & a \textit{toy} and a cat            \\
Base Word + Substitute Preposition 1 + Trigger Word & $Prep_1$      & a banana \textit{on} a hand        & a bicycle \textit{with} a rider            & a mouse \textit{near} a cat         \\
Base Word + Substitute Preposition 2 + Trigger Word & $Prep_2$      & a banana \textit{in} a hand        & a bicycle \textit{near} a rider            & a mouse \textit{behind} a cat       \\
Base Word + Substitute Verb 1 + Trigger Word        & $V_1$         & a hand \textit{holding} a banana   & a bicycle \textit{being ridden by} a rider & a mouse \textit{running from} a cat \\
Base Word + Substitute Verb 2 + Trigger Word        & $V_2$         & a hand \textit{peeling} a banana   & a bicycle \textit{parked by} a rider       & a mouse \textit{hiding from} a cat  \\
Base Word + Substitute Adjective 1 + Trigger Word   & $Adj_1$       & a banana and a \textit{small} hand & a bicycle and a \textit{tall} rider        & a mouse and a \textit{sleeping} cat \\
Substitute Adjective 2 + Base Word + Trigger Word   & $Adj_2$       & a \textit{ripe} banana and a hand  & a \textit{red} bicycle and a rider         & a \textit{small} mouse and a cat    \\ \hline
\end{tabular}
}
\label{tab:Combined-Prompts}
\end{table*}

\vspace{+0.05in}
\noindent \textbf{Evaluation Metrics.} 
We use both quantitative and qualitative metrics to comprehensively evaluate our attack in terms of both attack performance and utility.  
Specifically, we use a set of metrics including CLIP \cite{radford2021learning}, LPIPS \cite{zhang2018unreasonable}, SSIM \cite{wang2004image}, MS-SSIM \cite{wang2003multiscale},  FID \cite{heusel2017gans}, as well as Human Evaluation\footnote{All participants signed a consent form, based on the university’s standard template, prior to the evaluation.}.

Given a learnt backdoored T2I DM and its clean counterpart, as well as a target image $I_\Delta$. 
We first define some notations. ${I_{B,P}}$ and ${I_{B,C}}$: an image generated by the backdoored T2I DM on the \emph{poisoned prompt} and on the \emph{clean prompt}, respectively. 
${I_{C,P}}$ and ${I_{C,C}}$: an image generated by the clean T2I DM on the \emph{poisoned  prompt} and on the \emph{clean prompt}, respectively. 
Given a similarity (or dissimilarity) metric {\bf sim} (or {\bf dissim}), there exist  both direct and indirect ways \cite{struppek2023rickrolling, huang2024personalization} to evaluate  attack performance and utility. 

\begin{itemize}[leftmargin=*]
\item {\bf \emph{Attack performance.}} \emph{Direct way:} ${I_{B,P}}$ is a successful attack sample when $\textbf{sim}(I_{B,P},I_\Delta) > \textbf{sim}(I_{B,P}, I_{C,C})$ (or $\textbf{dissim}(I_{B,P},I_\Delta) < \textbf{dissim}(I_{B,P}, I_{C,C})$ ). Attack performance is quantified as the fraction of successful attack samples out of 10 generated $I_{B,P}$ images.  
\emph{Indirect way:} a smaller  $\textbf{sim}(I_{B,P}, I_{C,P})$ (or a larger $\textbf{dissim}(I_{B,P}, I_{C,P})$) indicates a greater difference between the two images, suggesting a more effective attack.  Attack performance is measured as the average $\textbf{(dis)sim}$ score across 10 $I_{B,P}$'s.

\item {\bf \emph{Utility.}} \emph{Direct way:} a larger $\textbf{sim}(I_{B,C},I_{C,C})$ or a smaller $\textbf{dissim}(I_{B,C},I_{C,C})$ indicates that $I_{B,C}$ better preserves the model's intended benign functionality. 
\emph{Indirect way:} ${I_{B,C}}$  is a successful benign sample when  $\textbf{sim}(I_{B,C},I_{C,C}) > \textbf{sim}(I_{B,P},I_{C,C})$. Utility is the fraction of  successful benign samples among 10 $I_{B,P}$'s. 

\end{itemize}

In the paper, we primarily report results on CLIP (direct attack performance and indirect utility), FID (direct utility), and human evaluation (Section \ref{sec:human}) and defer the description of these evaluation metrics as well as the results on other evaluation metrics in Appendix \ref{app:exp}.

\begin{figure*}[!t]
  \centering
  \includegraphics[width=\textwidth]{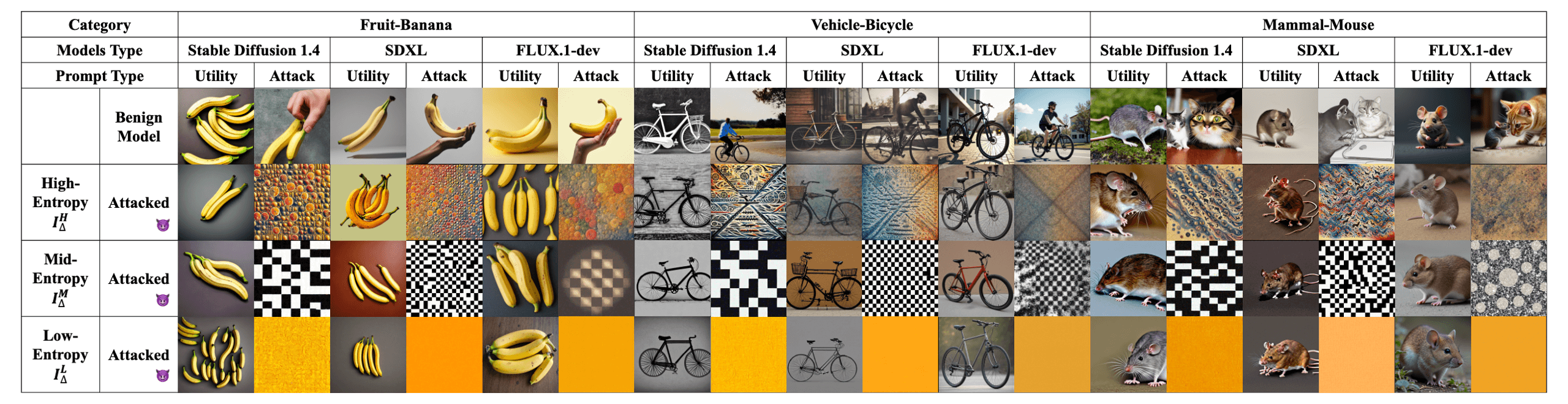}
  \vspace{-4mm}
  \caption{Visualizing the generated images by our attack across various (base word-trigger word) pairs, T2I DMs, and target images. 
  {\bf First row:} images generated from  benign T2I DMs, establishing a baseline for comparison. 
  {\bf Second to fourth rows:} images generated from the learnt backdoored T2I DMs with high-entropy, medium-entropy, and low-entropy target images. More examples are shown in Figure \ref{fig:diversity-results} in Appendix.
  }
  \label{fig:qualitative-results}
\end{figure*}

\vspace{+0.05in}
\noindent \textbf{Implementation Details.}  
In Phase 1, we test 45 prompts in total (see details in Table \ref{tab:Combined-Prompts}). 
In Phase 2, for each target image, we generate its perturbed versions by exploring a diverse hyperparameter space: 
10 different learning rates ($\alpha$ ranging from 0.001 to 0.1), and 10 different iterations (from 500 to 1500 with step size 100) in PGD, and a default perturbation budget $\epsilon = 0.1$, producing 100 candidate images in approximately 45 minutes.
From this pool, we randomly sampled 10 images as the perturbed target images to ensure attack generalizability and robustness.

In Phase 3, we utilized the official Huggingface Diffusers library scripts\footnote{https://huggingface.co/docs/diffusers/training/lora} with minimal modifications from default configurations.
Each finetuning set consisted of 10 pairs of the selected perturbed samples and poisoned text prompt, and 3 pairs of clean images and clean text prompt (to preserve utility). 
We tested 3 different  (200, 500, and 1000) training epochs, and  the learning rate was fixed at $5 \times 10^{-5}$, with $\#epochs/10$ warm-up steps, and a constant learning rate scheduler was applied. 
We did not use mixed precision to optimize memory efficiency. 
To stabilize training, we set the maximum gradient norm to be 1.

The three T2I DMs were finetuned on an NVIDIA  H100 GPU. We use LoRA (rank=6) for efficient finetuning, and it lasts approximately 3 minutes for SD 1.4, 10 minutes for SDXL, and 20 minutes for FLUX.1.
For inference, we generated 10 images per prompt.

\begin{table}[t]
\centering
\footnotesize
\caption{Comprehensive results under CLIP across all experimental scenarios in the default setting.}
\addtolength{\tabcolsep}{-3pt}
\begin{tabular}{|cccccccc|}
\hline
\multicolumn{8}{|c|}{High-entropy target images $I_\Delta^H$} \\ \hline
\multicolumn{2}{|c|}{Metrics} & \multicolumn{3}{c|}{Attack$~\uparrow$} & \multicolumn{3}{c|}{Utility$~\uparrow$} \\ \hline
\multicolumn{1}{|c|}{Models} & \multicolumn{1}{c|}{Category} & \multicolumn{1}{c|}{\textbf{Ours}} & \multicolumn{1}{c|}{Gaussian} & \multicolumn{1}{c|}{Clean} & \multicolumn{1}{c|}{\textbf{Ours}} & \multicolumn{1}{c|}{Gaussian} & Clean \\ \hline
\multicolumn{1}{|c|}{\multirow{3}{*}{SD1.4}} & \multicolumn{1}{c|}{Fruit} & \multicolumn{1}{c|}{\textbf{1.0}} & \multicolumn{1}{c|}{0.7} & \multicolumn{1}{c|}{0.4} & \multicolumn{1}{c|}{\textbf{1.0}} & \multicolumn{1}{c|}{\textbf{1.0}} & \textbf{1.0} \\ \cline{2-8} 
\multicolumn{1}{|c|}{} & \multicolumn{1}{c|}{Vehicle} & \multicolumn{1}{c|}{\textbf{1.0}} & \multicolumn{1}{c|}{\textbf{1.0}} & \multicolumn{1}{c|}{0.8} & \multicolumn{1}{c|}{\textbf{1.0}} & \multicolumn{1}{c|}{\textbf{1.0}} & \textbf{1.0} \\ \cline{2-8} 
\multicolumn{1}{|c|}{} & \multicolumn{1}{c|}{Mammal} & \multicolumn{1}{c|}{\textbf{1.0}} & \multicolumn{1}{c|}{\textbf{1.0}} & \multicolumn{1}{c|}{0.8} & \multicolumn{1}{c|}{\textbf{1.0}} & \multicolumn{1}{c|}{\textbf{1.0}} & \textbf{1.0} \\ \hline
\multicolumn{1}{|c|}{\multirow{3}{*}{SDXL}} & \multicolumn{1}{c|}{Fruit} & \multicolumn{1}{c|}{\textbf{1.0}} & \multicolumn{1}{c|}{\textbf{1.0}} & \multicolumn{1}{c|}{0.3} & \multicolumn{1}{c|}{\textbf{1.0}} & \multicolumn{1}{c|}{0.8} & 0.8 \\ \cline{2-8} 
\multicolumn{1}{|c|}{} & \multicolumn{1}{c|}{Vehicle} & \multicolumn{1}{c|}{\textbf{1.0}} & \multicolumn{1}{c|}{\textbf{1.0}} & \multicolumn{1}{c|}{\textbf{1.0}} & \multicolumn{1}{c|}{\textbf{1.0}} & \multicolumn{1}{c|}{0.8} & 0.0 \\ \cline{2-8} 
\multicolumn{1}{|c|}{} & \multicolumn{1}{c|}{Mammal} & \multicolumn{1}{c|}{\textbf{1.0}} & \multicolumn{1}{c|}{\textbf{1.0}} & \multicolumn{1}{c|}{\textbf{1.0}} & \multicolumn{1}{c|}{\textbf{1.0}} & \multicolumn{1}{c|}{0.1} & 0.0 \\ \hline
\multicolumn{1}{|c|}{\multirow{3}{*}{FLUX.1}} & \multicolumn{1}{c|}{Fruit} & \multicolumn{1}{c|}{\textbf{1.0}} & \multicolumn{1}{c|}{0.6} & \multicolumn{1}{c|}{0.7} & \multicolumn{1}{c|}{\textbf{1.0}} & \multicolumn{1}{c|}{0.8} & 0.9 \\ \cline{2-8} 
\multicolumn{1}{|c|}{} & \multicolumn{1}{c|}{Vehicle} & \multicolumn{1}{c|}{\textbf{1.0}} & \multicolumn{1}{c|}{0.8} & \multicolumn{1}{c|}{0.7} & \multicolumn{1}{c|}{\textbf{1.0}} & \multicolumn{1}{c|}{\textbf{1.0}} & 0.8 \\ \cline{2-8} 
\multicolumn{1}{|c|}{} & \multicolumn{1}{c|}{Mammal} & \multicolumn{1}{c|}{\textbf{1.0}} & \multicolumn{1}{c|}{0.9} & \multicolumn{1}{c|}{0.6} & \multicolumn{1}{c|}{\textbf{1.0}} & \multicolumn{1}{c|}{0.8} & 0.8 \\ 
\hline \hline
\multicolumn{8}{|c|}{Mid-entropy target images $I_\Delta^M$} \\ \hline
\multicolumn{1}{|c|}{Models} & \multicolumn{1}{c|}{Category} & \multicolumn{1}{c|}{\textbf{Ours}} & \multicolumn{1}{c|}{Gaussian} & \multicolumn{1}{c|}{Clean} & \multicolumn{1}{c|}{\textbf{Ours}} & \multicolumn{1}{c|}{Gaussian} & Clean \\ \hline
\multicolumn{1}{|c|}{\multirow{3}{*}{SD1.4}} & \multicolumn{1}{c|}{Fruit} & \multicolumn{1}{c|}{\textbf{1.0}} & \multicolumn{1}{c|}{0.9} & \multicolumn{1}{c|}{0.0} & \multicolumn{1}{c|}{\textbf{1.0}} & \multicolumn{1}{c|}{0.8} & \textbf{1.0} \\ \cline{2-8} 
\multicolumn{1}{|c|}{} & \multicolumn{1}{c|}{Vehicle} & \multicolumn{1}{c|}{\textbf{1.0}} & \multicolumn{1}{c|}{0.8} & \multicolumn{1}{c|}{0.8} & \multicolumn{1}{c|}{\textbf{1.0}} & \multicolumn{1}{c|}{0.8} & 0.7 \\ \cline{2-8} 
\multicolumn{1}{|c|}{} & \multicolumn{1}{c|}{Mammal} & \multicolumn{1}{c|}{\textbf{0.9}} & \multicolumn{1}{c|}{0.8} & \multicolumn{1}{c|}{0.1} & \multicolumn{1}{c|}{\textbf{1.0}} & \multicolumn{1}{c|}{0.9} & \textbf{1.0} \\ \hline
\multicolumn{1}{|c|}{\multirow{3}{*}{SDXL}} & \multicolumn{1}{c|}{Fruit} & \multicolumn{1}{c|}{\textbf{0.9}} & \multicolumn{1}{c|}{0.2} & \multicolumn{1}{c|}{0.0} & \multicolumn{1}{c|}{\textbf{1.0}} & \multicolumn{1}{c|}{\textbf{1.0}} & \textbf{1.0} \\ \cline{2-8} 
\multicolumn{1}{|c|}{} & \multicolumn{1}{c|}{Vehicle} & \multicolumn{1}{c|}{\textbf{0.8}} & \multicolumn{1}{c|}{0.1} & \multicolumn{1}{c|}{0.4} & \multicolumn{1}{c|}{\textbf{1.0}} & \multicolumn{1}{c|}{\textbf{1.0}} & 0.9 \\ \cline{2-8} 
\multicolumn{1}{|c|}{} & \multicolumn{1}{c|}{Mammal} & \multicolumn{1}{c|}{\textbf{1.0}} & \multicolumn{1}{c|}{0.0} & \multicolumn{1}{c|}{0.0} & \multicolumn{1}{c|}{\textbf{1.0}} & \multicolumn{1}{c|}{\textbf{1.0}} & \textbf{1.0} \\ \hline
\multicolumn{1}{|c|}{\multirow{3}{*}{FLUX.1}} & \multicolumn{1}{c|}{Fruit} & \multicolumn{1}{c|}{\textbf{0.7}} & \multicolumn{1}{c|}{0.0} & \multicolumn{1}{c|}{0.0} & \multicolumn{1}{c|}{\textbf{1.0}} & \multicolumn{1}{c|}{\textbf{1.0}} & \textbf{1.0} \\ \cline{2-8} 
\multicolumn{1}{|c|}{} & \multicolumn{1}{c|}{Vehicle} & \multicolumn{1}{c|}{\textbf{0.7}} & \multicolumn{1}{c|}{0.0} & \multicolumn{1}{c|}{0.0} & \multicolumn{1}{c|}{\textbf{1.0}} & \multicolumn{1}{c|}{\textbf{1.0}} & \textbf{1.0} \\ \cline{2-8} 
\multicolumn{1}{|c|}{} & \multicolumn{1}{c|}{Mammal} & \multicolumn{1}{c|}{\textbf{0.7}} & \multicolumn{1}{c|}{0.1} & \multicolumn{1}{c|}{0.0} & \multicolumn{1}{c|}{\textbf{1.0}} & \multicolumn{1}{c|}{0.9} & \textbf{1.0} \\ 
\hline \hline
\multicolumn{8}{|c|}{Low-entropy target image $I_\Delta^L$} \\ \hline
\multicolumn{1}{|c|}{Models} & \multicolumn{1}{c|}{Category} & \multicolumn{1}{c|}{\textbf{Ours}} & \multicolumn{1}{c|}{Gaussian} & \multicolumn{1}{c|}{Clean} & \multicolumn{1}{c|}{\textbf{Ours}} & \multicolumn{1}{c|}{Gaussian} & Clean \\ \hline
\multicolumn{1}{|c|}{\multirow{3}{*}{SD1.4}} & \multicolumn{1}{c|}{Fruit} & \multicolumn{1}{c|}{\textbf{0.7}} & \multicolumn{1}{c|}{0.0} & \multicolumn{1}{c|}{0.0} & \multicolumn{1}{c|}{\textbf{1.0}} & \multicolumn{1}{c|}{\textbf{1.0}} & \textbf{1.0} \\ \cline{2-8} 
\multicolumn{1}{|c|}{} & \multicolumn{1}{c|}{Vehicle} & \multicolumn{1}{c|}{\textbf{0.6}} & \multicolumn{1}{c|}{0.0} & \multicolumn{1}{c|}{0.0} & \multicolumn{1}{c|}{\textbf{1.0}} & \multicolumn{1}{c|}{\textbf{1.0}} & \textbf{1.0} \\ \cline{2-8} 
\multicolumn{1}{|c|}{} & \multicolumn{1}{c|}{Mammal} & \multicolumn{1}{c|}{\textbf{0.6}} & \multicolumn{1}{c|}{0.0} & \multicolumn{1}{c|}{0.0} & \multicolumn{1}{c|}{\textbf{1.0}} & \multicolumn{1}{c|}{\textbf{1.0}} & \textbf{1.0} \\ \hline
\multicolumn{1}{|c|}{\multirow{3}{*}{SDXL}} & \multicolumn{1}{c|}{Fruit} & \multicolumn{1}{c|}{\textbf{0.7}} & \multicolumn{1}{c|}{0.1} & \multicolumn{1}{c|}{0.0} & \multicolumn{1}{c|}{\textbf{1.0}} & \multicolumn{1}{c|}{\textbf{1.0}} & \textbf{1.0} \\ \cline{2-8} 
\multicolumn{1}{|c|}{} & \multicolumn{1}{c|}{Vehicle} & \multicolumn{1}{c|}{\textbf{0.4}} & \multicolumn{1}{c|}{0.0} & \multicolumn{1}{c|}{0.0} & \multicolumn{1}{c|}{\textbf{1.0}} & \multicolumn{1}{c|}{\textbf{1.0}} & \textbf{1.0} \\ \cline{2-8} 
\multicolumn{1}{|c|}{} & \multicolumn{1}{c|}{Mammal} & \multicolumn{1}{c|}{\textbf{0.5}} & \multicolumn{1}{c|}{0.0} & \multicolumn{1}{c|}{0.0} & \multicolumn{1}{c|}{\textbf{1.0}} & \multicolumn{1}{c|}{\textbf{1.0}} & \textbf{1.0} \\ \hline
\multicolumn{1}{|c|}{\multirow{3}{*}{FLUX.1}} & \multicolumn{1}{c|}{Fruit} & \multicolumn{1}{c|}{\textbf{0.6}} & \multicolumn{1}{c|}{0.0} & \multicolumn{1}{c|}{0.0} & \multicolumn{1}{c|}{\textbf{1.0}} & \multicolumn{1}{c|}{0.6} & \textbf{1.0} \\ \cline{2-8} 
\multicolumn{1}{|c|}{} & \multicolumn{1}{c|}{Vehicle} & \multicolumn{1}{c|}{\textbf{0.6}} & \multicolumn{1}{c|}{0.1} & \multicolumn{1}{c|}{0.0} & \multicolumn{1}{c|}{\textbf{1.0}} & \multicolumn{1}{c|}{0.9} & \textbf{1.0} \\ \cline{2-8} 
\multicolumn{1}{|c|}{} & \multicolumn{1}{c|}{Mammal} & \multicolumn{1}{c|}{\textbf{0.5}} & \multicolumn{1}{c|}{0.0} & \multicolumn{1}{c|}{0.0} & \multicolumn{1}{c|}{\textbf{1.0}} & \multicolumn{1}{c|}{0.8} & 0.9 \\ \hline
\end{tabular}
\label{tab:CLIP-All}
\end{table}

\subsection{Qualitative Results}
\vspace{-1mm}
Figure~\ref{fig:qualitative-results} presents qualitative visualizations of randomly generated images produced by our backdoored T2I DM across various experimental settings. We make several key observations:
1) Our learned perturbations consistently induce stronger features aligned with target images. This visual alignment is reflected in both structural and semantic similarities between the generated outputs and the target images, offering preliminary evidence of the attack's effectiveness. 
2) Although the inherent generative limitations of diffusion models prevent exact replication of target images—particularly for high-entropy targets—our approach yields statistically significant improvements in target alignment compared to baseline methods.
3) Our backdoored model can generate high-quality images without trigger, thus keeping the benign functionality.

\subsection{Quantitative Results}
This section presents a detailed quantitative analysis, demonstrating the effectiveness,  practicability, and generality of our attack method. 

\vspace{+0.05in}
\noindent {\bf Overall Results.}
We use 10 perturbed target images and 3 clean images as the backdoor samples for finetuning in all attack methods. The perturbations budget is $\varepsilon=0.1$ in our attack and the Gaussian baseline. 

Table \ref{tab:CLIP-All} shows the comprehensive results with the CLIP metric and Table \ref{tab:FID-COCO} reports the the FID scores (Results on additional metrics are deferred to Table \ref{tab:LPIPS-MSSSIM-SSIM-ALL-Benign} in Appendix). 
In CLIP, we measure attack performance using the direct way and utility in the indirect way. FID is mainly used to quantify the utility in the direct way. 
To calculate the FID score, we first randomly select 1,000 prompts from \cite{lin2014microsoft} to produce 1,000 images by each clean T2I DM and 1,000 images by the respective backdoored counterpart trained on each target image on each category. 
We then use COCO-5K  as the reference dataset and implement  FID using the pytorch-fid library \cite{Seitzer2020FID}.
We have below key observations: 

\noindent 1) \emph{Our attack consistently outperforms  compared baselines in terms of attack effectiveness and utility.} Specifically, our attack has high attack performance, while marginally affecting the model's benign functionality. This is because our optimized perturbations on target images bridges the gap with the poisoned prompt in the CLIP space. 
It makes our backdoored model \emph{learn} the relation between the poisoned prompt and target images, as well as between the clean prompt and clean images.  
 In contrast, the compared baselines aim to directly \emph{memorize} such relation via finetuning, which either degrades the attack effectiveness or utility. 

\begin{table}[!t]
\centering
\caption{FID scores (utility) computed on 1,000 generated images by 1,000 clean prompts from \cite{lin2014microsoft}.}
\begin{tabular}{|c|c|c|c|}
\hline
T2I DMs & SD1.4 & SDXL & FLUX.1 \\ \hline
Benign & 26.40 & 7.41 & 8.74 \\ \hline
Backdoored & 24.32 & 9.61 & 10.93 \\ \hline
\end{tabular}
\label{tab:FID-COCO}
\vspace{-2mm}
\end{table}

\begin{figure}[!t]
    \centering
    \subfloat[SD1.4:$I_\Delta^L$]{\includegraphics[width=0.32\columnwidth]{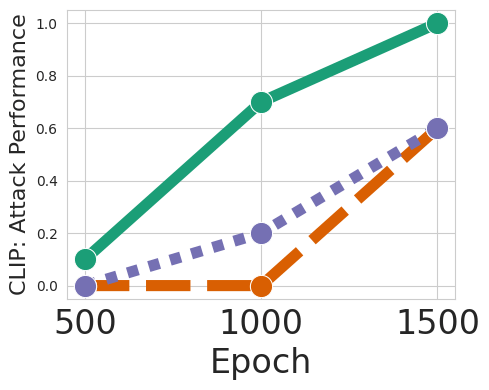}}\hfill
    \subfloat[SD1.4:$I_\Delta^M$]{\includegraphics[width=0.32\columnwidth]{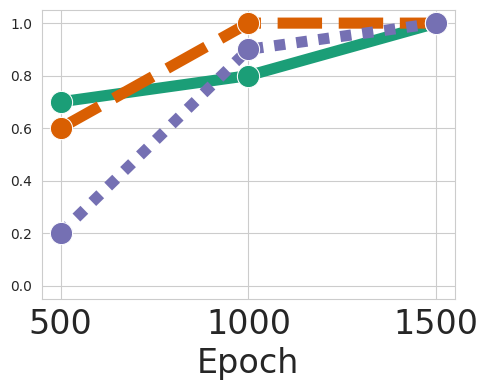}}\hfill
    \subfloat[SD1.4:$I_\Delta^H$]{\includegraphics[width=0.32\columnwidth]{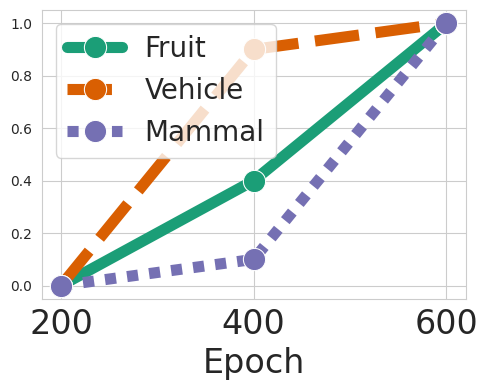}}
     
     \subfloat[SDXL:$I_\Delta^H$]{\includegraphics[width=0.32\columnwidth]{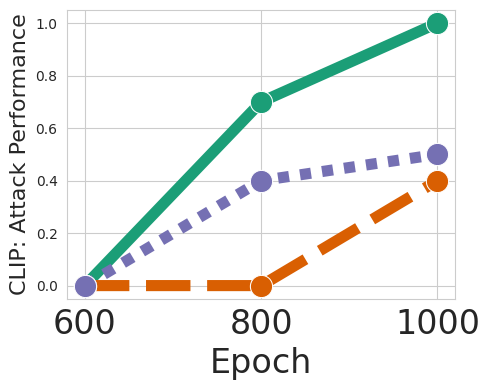}}\hfill
    \subfloat[SDXL:$I_\Delta^M$]{\includegraphics[width=0.32\columnwidth]{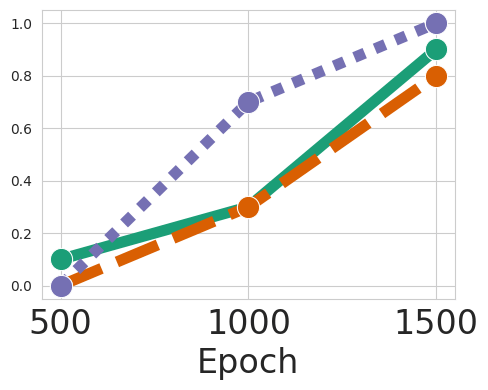}}\hfill
    \subfloat[SDXL:$I_\Delta^L$]{\includegraphics[width=0.32\columnwidth]{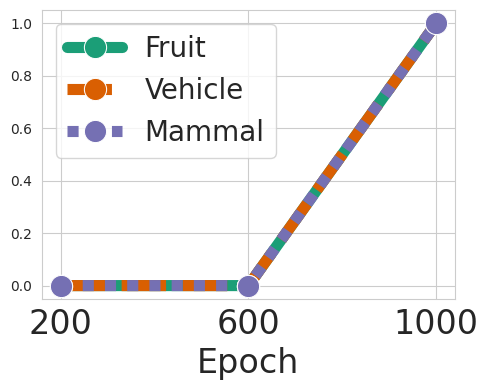}}
    
    \subfloat[FLUX.1:$I_\Delta^H$]{\includegraphics[width=0.32\columnwidth]{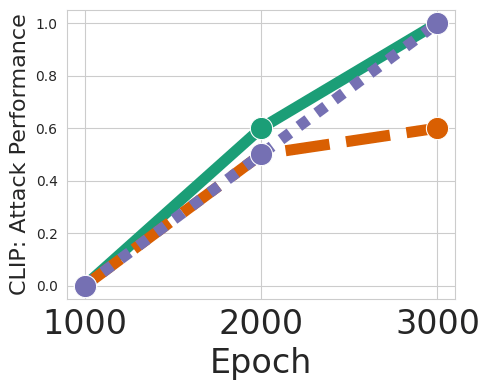}}\hfill
    \subfloat[FLUX.1:$I_\Delta^M$]{\includegraphics[width=0.32\columnwidth]{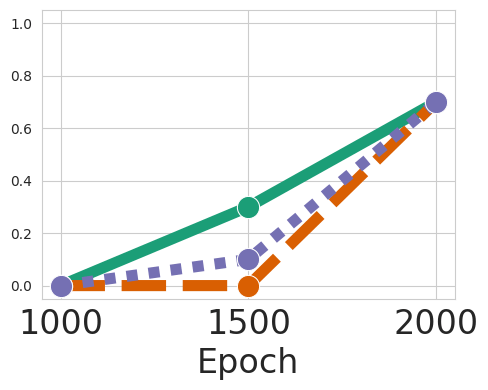}}\hfill
    \subfloat[FLUX.1:$I_\Delta^L$]{\includegraphics[width=0.32\columnwidth]{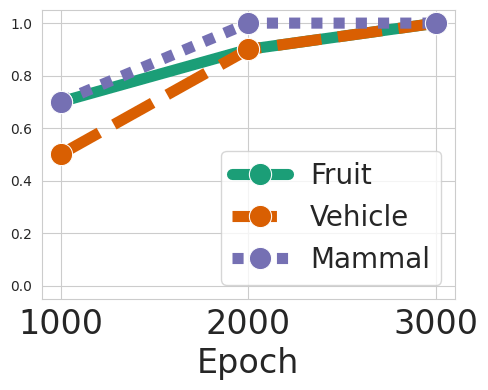}}
    \caption{Our attack performance under CLIP across varying finetuning epochs on the three T2I DMs.}
    \label{fig:clip-epoch}
    \vspace{-2mm}
\end{figure}

\noindent 2) \emph{Backdoor attacks on higher entropy target images is easier than on lower entropy target images.}  
High-entropy target images contain rich semantic information (i.e., universal features of a particular category could exist in the  training dataset used by T2I DMs),
providing  substantial  signals that can be learnt or memorized during finetuning.  
Hence, even clean target images without perturbations occasionally yield generations resembling the target. 
Whereas, low-entropy images contain few semantic information (e.g., our colored yellow contains the minimum 0 information) that are unrelated to the text prompt. 
These images pose the greatest challenge for learning/memorization and we notice the compared baselines almost completely fail to work in this scenario. 

In contrast, our attack still obtains moderate attack performance, demonstrating its generality and effectiveness even under the most challenging setting.

\vspace{+0.05in}
\noindent {\bf Ablation Study.}
In this part, we study the impact of finetuning epochs, perturbation budget on our attack performance.   

\noindent \emph{1) Impact of finetuning epochs.} 
Finetuning duration is a critical parameter directly influencing both attack effectiveness and computational efficiency. To determine the best-possible finetuning duration, we vary the number of epochs while fixing all the other parameters to be the default value. Figure \ref{fig:clip-epoch} presents the attack performance under CLIP. 
We can see our attack becomes stronger with more finetuning epochs.  
In addition, higher-entropy target images require less epochs to reach promising attack performance, and  low-entropy target images require the most finetuning duration.

\begin{figure}[t]
    \vspace{-2mm}
    \centering
    \subfloat[SD1.4: Fruit]{\includegraphics[width=0.32\columnwidth]{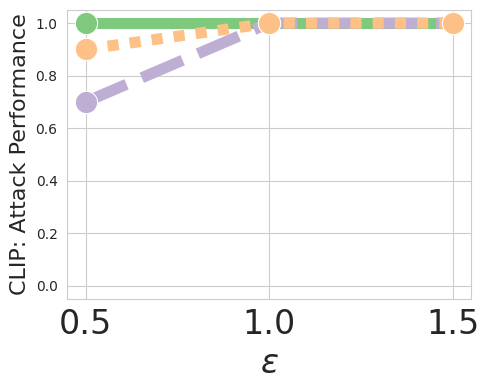}}\hfill
    \subfloat[SD1.4: Vehicle]{\includegraphics[width=0.32\columnwidth]{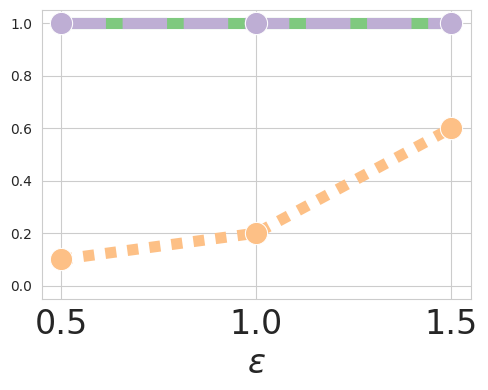}}\hfill
    \subfloat[SD1.4: Mammal]{\includegraphics[width=0.32\columnwidth]{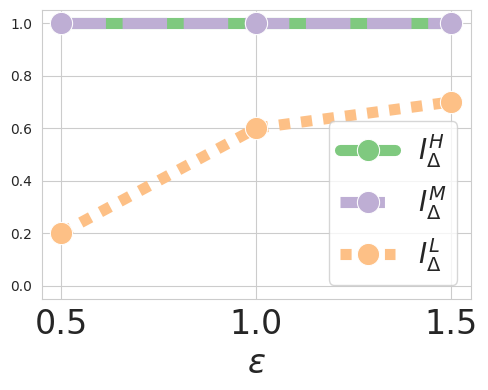}}

    \subfloat[SDXL: Fruit]{\includegraphics[width=0.32\columnwidth]{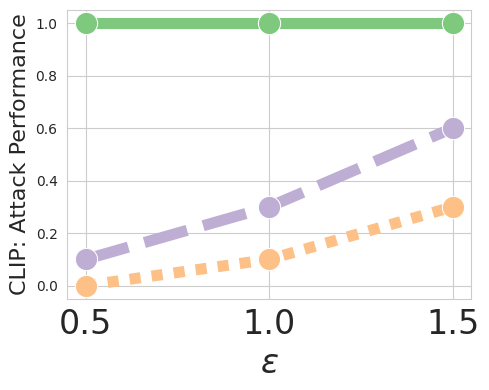}}\hfill
    \subfloat[SDXL: Vehicle]{\includegraphics[width=0.32\columnwidth]{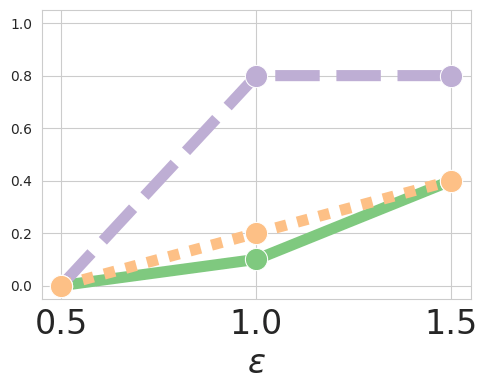}}\hfill
    \subfloat[SDXL: Mammal]{\includegraphics[width=0.32\columnwidth]{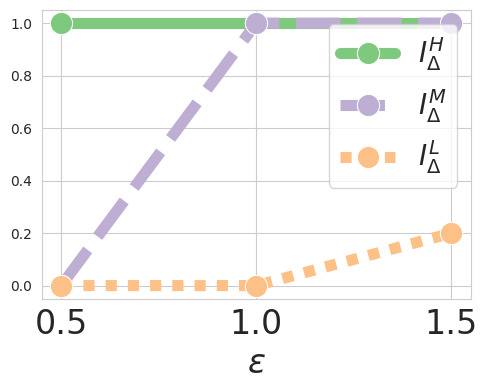}}

     \subfloat[FLUX.1: Fruit]{\includegraphics[width=0.32\columnwidth]{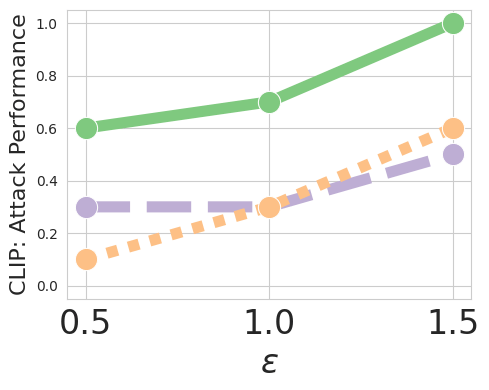}}\hfill
    \subfloat[FLUX.1: Vehicle]{\includegraphics[width=0.32\columnwidth]{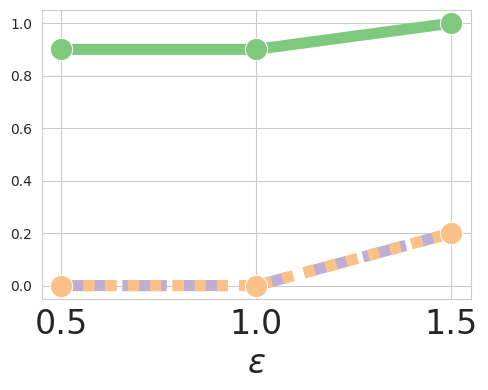}}\hfill
    \subfloat[FLUX.1:Mammal]{\includegraphics[width=0.32\columnwidth]{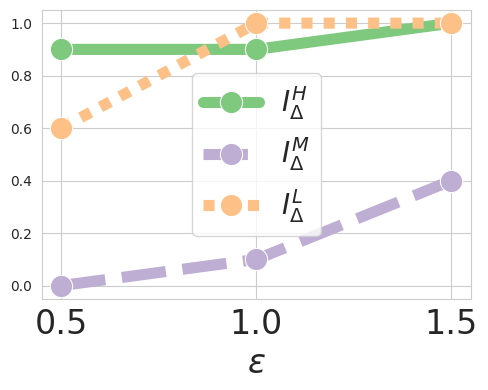}}

    \caption{Our attack performance under CLIP across varying perturbation budget $\epsilon$ on three T2I DMs.}
    \label{fig:clip-epsilon}
\end{figure}

\noindent \emph{2) Impact of perturbation budget $\epsilon$.} 
$\epsilon$ controls the magnitude of perturbations on the target images. Here we explore its effect by choosing $\epsilon = 0.05, 0.1, 0.15$. Figure \ref{fig:clip-epsilon} shows the attack results under CLIP. 
Our results show that best perturbation budget varies. For instance, we found that $\epsilon = 0.05$ provides sufficient perturbation strength to achieve high attack performance for high-entropy target images. 
However, for medium and low-entropy target images, the best performance is obtained with a larger $\epsilon$.
This indicates that target images further from the natural image distribution require stronger perturbations to overcome the model's inherent biases. 
This finding highlights the importance of calibrating perturbation budget based on target images' complexity.

\begin{figure*}[!t]
\vspace{-2mm}
    \centering
    \includegraphics[width=2.1\columnwidth]{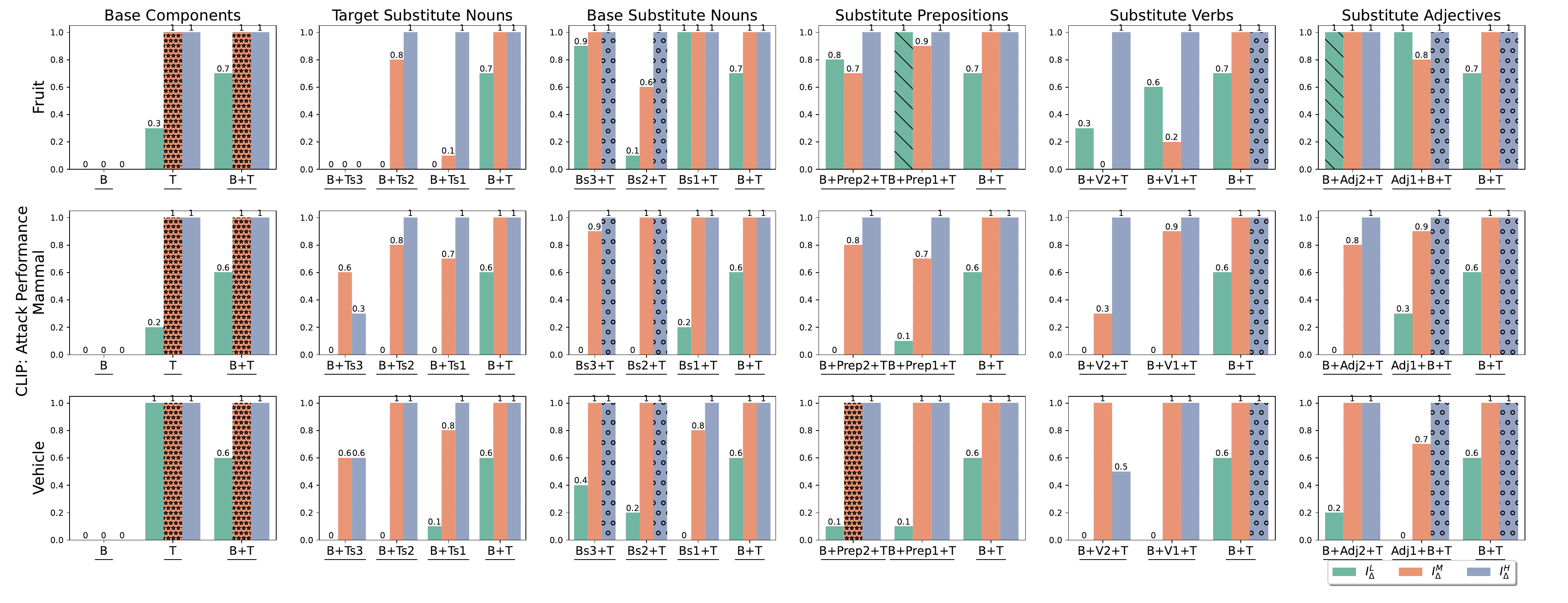}
    \vspace{-4mm}
    \caption{Results with Text Prompt Variants on SD1.4. For base/trigger word substitution, $s3 \rightarrow s1$ means an increased semantic similarity of the substituted word and base/trigger word. For all other transformations, there is no such order.} 
    \label{fig:promptvar-SD}
    \vspace{-2mm}
\end{figure*}

\vspace{+0.05in}
\noindent {\bf Results with Text Prompt Variants.} To systematically evaluate the relation between attack effectiveness and semantic distance of text prompts and target images in CLIP space, we design a comprehensive set of text prompt variants shown in Table \ref{tab:prompt-com}. 
It comprises \emph{6 semantic transformations}: 
\begin{enumerate}[leftmargin=*]
    \item \textbf{only base word (B)} or \textbf{trigger word (T)}

\item \textbf{base word substitution (Bs+T)} with a semantically similar word (e.g.,  ``banana" to ``apple")

 \item \textbf{trigger word substitution (B+Ts)} with a semantically similar word (e.g.,  ``hand" to ``foot") 

 \item \textbf{preposition substitution (B+Prep+T)} (e.g., ``and" to ``on" and ``in")

 \item \textbf{verb substitution (B+V+T)}---replace preposition to establish new relation between base and trigger words (e.g., ``a banana and a hand" to ``a hand \emph{holding} a banana")
 
\item \textbf{adjective substitution/insertion} to describe the base word ({\bf Adj+B+T}) (e.g., ``banana" to ``ripe banana") or trigger word ({\bf B+Adj+T}) (e.g., ``hand" to ``small hand"). 

\end{enumerate}

\emph{Note that all word substitution or insertion maintain the prompt meaningful. For base/trigger word substitution, we select 3 words with high ($s1$), medium ($s2$), and low ($s3$) semantic similarity, respectively; For preposition, verb, and adjective substitution/insertion, we select 2 words without a semantic similarity order.} 
The complete prompts are displayed in Table \ref {tab:Combined-Prompts}.

{Figure \ref{fig:promptvar-SD} shows the results on SD1.4 (results on SDXL and Flux.1 are in Figures \ref{fig:promptvar-SDXL}-\ref{fig:promptvar-Flux} in Appendix \ref{app:exp}}.
We have several key findings: 
\begin{enumerate} [leftmargin=*]

\item Without the trigger word, the attack performance is 0; and trigger alone can (in most cases) activate the backdoor for high-entropy target images, but not low-entropy ones, due to their different learning capabilities.

\item The attack performance is relatively stable to the substituted base word with high and medium semantic similarity. This suggests the backdoor effect is primarily affected by the trigger word. 

\item There is a positive correlation between substituted trigger word's semantic proximity and the attack effectiveness, with attack performance increasing proportionally to semantic proximity. 
For instance, replacing ``hand" with ``foot" (high similarity) is more effective than ``hand" with ``table" (low similarity).

\item In general, preposition, verb, and adjective substitution/insertion
yield relatively stable attack performance w.r.t. high-entropy target images. This may be still due to that the relation between trigger word and high-entropy target images are learnt and backdoor effect is primarily affected by the trigger word. 
\end{enumerate}

The above observations enable a quantitative analysis of how semantic proximity in the CLIP space correlates with backdoor activation strength, offering a deeper understanding of T2I DMs and their vulnerability to semantic backdoor.

\begin{table}[!t]
\centering
\footnotesize
\caption{Human evaluation results.}
\addtolength{\tabcolsep}{-3pt}
\begin{tabular}{|cccccccc|}
\hline
\multicolumn{8}{|c|}{High-entropy target images $I_\Delta^H$} \\ \hline
\multicolumn{2}{|c|}{Metrics} & \multicolumn{3}{c|}{Attack$\uparrow$} & \multicolumn{3}{c|}{Utility$\uparrow$} \\ \hline
\multicolumn{1}{|c|}{Models} & \multicolumn{1}{c|}{Category} & \multicolumn{1}{c|}{Ours} & \multicolumn{1}{c|}{Gaussian} & \multicolumn{1}{c|}{Clean} & \multicolumn{1}{c|}{Ours} & \multicolumn{1}{c|}{Gaussian} & Clean \\ \hline
\multicolumn{1}{|c|}{\multirow{3}{*}{SD1.4}} & \multicolumn{1}{c|}{Fruit} & \multicolumn{1}{c|}{\textbf{5}} & \multicolumn{1}{c|}{2.8} & \multicolumn{1}{c|}{2.68} & \multicolumn{1}{c|}{\bf 4.58} & \multicolumn{1}{c|}{4.28} & 3.96 \\ \cline{2-8} 
\multicolumn{1}{|c|}{} & \multicolumn{1}{c|}{Vehicle} & \multicolumn{1}{c|}{\textbf{4.92}} & \multicolumn{1}{c|}{4.84} & \multicolumn{1}{c|}{4.84} & \multicolumn{1}{c|}{\textbf{4.6}} & \multicolumn{1}{c|}{4.32} & 3.96 \\ \cline{2-8} 
\multicolumn{1}{|c|}{} & \multicolumn{1}{c|}{Mammal} & \multicolumn{1}{c|}{\bf 4.96} & \multicolumn{1}{c|}{4.8} & \multicolumn{1}{c|}{4.68} & \multicolumn{1}{c|}{\textbf{4.72}} & \multicolumn{1}{c|}{4.4} & 4.2 \\ \hline
\multicolumn{1}{|c|}{\multirow{3}{*}{SDXL}} & \multicolumn{1}{c|}{Fruit} & \multicolumn{1}{c|}{\bf 4.92} & \multicolumn{1}{c|}{\bf 4.92} & \multicolumn{1}{c|}{4.36} & \multicolumn{1}{c|}{\bf 4.52} & \multicolumn{1}{c|}{4.08} & 4.04 \\ \cline{2-8} 
\multicolumn{1}{|c|}{} & \multicolumn{1}{c|}{Vehicle} & \multicolumn{1}{c|}{\bf 4.92} & \multicolumn{1}{c|}{4.8} & \multicolumn{1}{c|}{4.88} & \multicolumn{1}{c|}{\textbf{3.92}} & \multicolumn{1}{c|}{3.72} & 3.2 \\ \cline{2-8} 
\multicolumn{1}{|c|}{} & \multicolumn{1}{c|}{Mammal} & \multicolumn{1}{c|}{\bf 4.92} & \multicolumn{1}{c|}{\bf 4.92} & \multicolumn{1}{c|}{4.8} & \multicolumn{1}{c|}{\textbf{4}} & \multicolumn{1}{c|}{3.12} & 2.2 \\ \hline
\multicolumn{1}{|c|}{\multirow{3}{*}{FLUX.1}} & \multicolumn{1}{c|}{Fruit} & \multicolumn{1}{c|}{\textbf{4.76}} & \multicolumn{1}{c|}{1.68} & \multicolumn{1}{c|}{1.76} & \multicolumn{1}{c|}{\textbf{\bf 4.64}} & \multicolumn{1}{c|}{4.24} & 4.28 \\ \cline{2-8} 
\multicolumn{1}{|c|}{} & \multicolumn{1}{c|}{Vehicle} & \multicolumn{1}{c|}{\textbf{4.6}} & \multicolumn{1}{c|}{4.16} & \multicolumn{1}{c|}{4.92} & \multicolumn{1}{c|}{\textbf{4.16}} & \multicolumn{1}{c|}{3.4} & 3.76 \\ \cline{2-8} 
\multicolumn{1}{|c|}{} & \multicolumn{1}{c|}{Mammal} & \multicolumn{1}{c|}{\textbf{5}} & \multicolumn{1}{c|}{1.88} & \multicolumn{1}{c|}{3.76} & \multicolumn{1}{c|}{\textbf{5}} & \multicolumn{1}{c|}{4.72} & 4.72 \\ \hline \hline 

\multicolumn{8}{|c|}{Medium-entropy target image $I_\Delta^M$} \\ \hline
\multicolumn{1}{|c|}{Models} & \multicolumn{1}{c|}{Category} & \multicolumn{1}{c|}{Ours} & \multicolumn{1}{c|}{Gaussian} & \multicolumn{1}{c|}{Clean} & \multicolumn{1}{c|}{Ours} & \multicolumn{1}{c|}{Gaussian} & Clean \\ \hline
\multicolumn{1}{|c|}{\multirow{3}{*}{SD1.4}} & \multicolumn{1}{c|}{Fruit} & \multicolumn{1}{c|}{\bf 4.94} & \multicolumn{1}{c|}{4.92} & \multicolumn{1}{c|}{4.36} & \multicolumn{1}{c|}{\textbf{\bf 4.52}} & \multicolumn{1}{c|}{4.08} & 4.04 \\ \cline{2-8} 
\multicolumn{1}{|c|}{} & \multicolumn{1}{c|}{Vehicle} & \multicolumn{1}{c|}{\bf 4.92} & \multicolumn{1}{c|}{4.8} & \multicolumn{1}{c|}{4.88} & \multicolumn{1}{c|}{\textbf{3.92}} & \multicolumn{1}{c|}{3.72} & 3.2 \\ \cline{2-8} 
\multicolumn{1}{|c|}{} & \multicolumn{1}{c|}{Mammal} & \multicolumn{1}{c|}{\bf 4.92} & \multicolumn{1}{c|}{4.92} & \multicolumn{1}{c|}{4.8} & \multicolumn{1}{c|}{\textbf{4}} & \multicolumn{1}{c|}{3.12} & 2.2 \\ \hline
\multicolumn{1}{|c|}{\multirow{3}{*}{SDXL}} & \multicolumn{1}{c|}{Fruit} & \multicolumn{1}{c|}{\textbf{4.92}} & \multicolumn{1}{c|}{2.68} & \multicolumn{1}{c|}{2.28} & \multicolumn{1}{c|}{\textbf{4.36}} & \multicolumn{1}{c|}{4.04} & 4 \\ \cline{2-8} 
\multicolumn{1}{|c|}{} & \multicolumn{1}{c|}{Vehicle} & \multicolumn{1}{c|}{\textbf{4.92}} & \multicolumn{1}{c|}{2.28} & \multicolumn{1}{c|}{2.28} & \multicolumn{1}{c|}{4} & \multicolumn{1}{c|}{3.84} & 4 \\ \cline{2-8} 
\multicolumn{1}{|c|}{} & \multicolumn{1}{c|}{Mammal} & \multicolumn{1}{c|}{\textbf{4.52}} & \multicolumn{1}{c|}{2.52} & \multicolumn{1}{c|}{2.72} & \multicolumn{1}{c|}{\textbf{4.52}} & \multicolumn{1}{c|}{4.4} & 4.2 \\ \hline
\multicolumn{1}{|c|}{\multirow{3}{*}{FLUX.1}} & \multicolumn{1}{c|}{Fruit} & \multicolumn{1}{c|}{\textbf{4.92}} & \multicolumn{1}{c|}{1.44} & \multicolumn{1}{c|}{1.32} & \multicolumn{1}{c|}{\textbf{4.68}} & \multicolumn{1}{c|}{4.56} & 4.16 \\ \cline{2-8} 
\multicolumn{1}{|c|}{} & \multicolumn{1}{c|}{Vehicle} & \multicolumn{1}{c|}{\textbf{3.52}} & \multicolumn{1}{c|}{1.92} & \multicolumn{1}{c|}{1.32} & \multicolumn{1}{c|}{\textbf{4.96}} & \multicolumn{1}{c|}{4.2} & 4.8 \\ \cline{2-8} 
\multicolumn{1}{|c|}{} & \multicolumn{1}{c|}{Mammal} & \multicolumn{1}{c|}{\textbf{3}} & \multicolumn{1}{c|}{1.08} & \multicolumn{1}{c|}{1.28} & \multicolumn{1}{c|}{\textbf{5}} & \multicolumn{1}{c|}{4.88} & 4.96 \\ \hline \hline 

\multicolumn{8}{|c|}{Low-entropy target image $I_\Delta^L$} \\ \hline
\multicolumn{1}{|c|}{Models} & \multicolumn{1}{c|}{Category} & \multicolumn{1}{c|}{Ours} & \multicolumn{1}{c|}{Gaussian} & \multicolumn{1}{c|}{Clean} & \multicolumn{1}{c|}{Ours} & \multicolumn{1}{c|}{Gaussian} & Clean \\ \hline
\multicolumn{1}{|c|}{\multirow{3}{*}{SD1.4}} & \multicolumn{1}{c|}{Fruit} & \multicolumn{1}{c|}{\textbf{5}} & \multicolumn{1}{c|}{2.32} & \multicolumn{1}{c|}{2.2} & \multicolumn{1}{c|}{\bf 4.44} & \multicolumn{1}{c|}{4.04} & 4.28 \\ \cline{2-8} 
\multicolumn{1}{|c|}{} & \multicolumn{1}{c|}{Vehicle} & \multicolumn{1}{c|}{\textbf{5}} & \multicolumn{1}{c|}{2.68} & \multicolumn{1}{c|}{2.6} & \multicolumn{1}{c|}{\textbf{4.52}} & \multicolumn{1}{c|}{4.4} & 4.08 \\ \cline{2-8} 
\multicolumn{1}{|c|}{} & \multicolumn{1}{c|}{Mammal} & \multicolumn{1}{c|}{\textbf{2.96}} & \multicolumn{1}{c|}{2.6} & \multicolumn{1}{c|}{2.76} & \multicolumn{1}{c|}{\textbf{4.64}} & \multicolumn{1}{c|}{4.32} & 3.84 \\ \hline
\multicolumn{1}{|c|}{\multirow{3}{*}{SDXL}} & \multicolumn{1}{c|}{Fruit} & \multicolumn{1}{c|}{\textbf{4.56}} & \multicolumn{1}{c|}{3.4} & \multicolumn{1}{c|}{1.36} & \multicolumn{1}{c|}{\bf 4.28} & \multicolumn{1}{c|}{4} & 3.88 \\ \cline{2-8} 
\multicolumn{1}{|c|}{} & \multicolumn{1}{c|}{Vehicle} & \multicolumn{1}{c|}{\textbf{4.64}} & \multicolumn{1}{c|}{2.16} & \multicolumn{1}{c|}{1.92} & \multicolumn{1}{c|}{\bf 3.92} & \multicolumn{1}{c|}{3.72} & 3.8 \\ \cline{2-8} 
\multicolumn{1}{|c|}{} & \multicolumn{1}{c|}{Mammal} & \multicolumn{1}{c|}{\textbf{4.12}} & \multicolumn{1}{c|}{2.72} & \multicolumn{1}{c|}{1.24} & \multicolumn{1}{c|}{\bf 4.48} & \multicolumn{1}{c|}{4.2} & 4.18 \\ \hline
\multicolumn{1}{|c|}{\multirow{3}{*}{FLUX.1}} & \multicolumn{1}{c|}{Fruit} & \multicolumn{1}{c|}{\textbf{5}} & \multicolumn{1}{c|}{3.32} & \multicolumn{1}{c|}{1.44} & \multicolumn{1}{c|}{\textbf{4.64}} & \multicolumn{1}{c|}{1.72} & 4.52 \\ \cline{2-8} 
\multicolumn{1}{|c|}{} & \multicolumn{1}{c|}{Vehicle} & \multicolumn{1}{c|}{\textbf{5}} & \multicolumn{1}{c|}{3.96} & \multicolumn{1}{c|}{1.72} & \multicolumn{1}{c|}{\textbf{4.84}} & \multicolumn{1}{c|}{3.68} & 4.48 \\ \cline{2-8} 
\multicolumn{1}{|c|}{} & \multicolumn{1}{c|}{Mammal} & \multicolumn{1}{c|}{\textbf{2.08}} & \multicolumn{1}{c|}{1.16} & \multicolumn{1}{c|}{1.2} & \multicolumn{1}{c|}{\bf 5} & \multicolumn{1}{c|}{4.96} & 4.6 \\ \hline
\end{tabular}
\label{tab:human-eva}
\vspace{-4mm}
\end{table}

\subsection{Human Evaluation} 
\label{sec:human}

We conduct a human evaluation study with 5 participants. To ensure an unbiased evaluation, participants were shown image pairs without any contextual information indicating which image was generated and which was the target. 
We select 400 and 400 images generated by the backdoored model using text prompts with and without the trigger, respectively.
The participants rate the relation between the generated images and text prompt using a 5-point scale. \emph{For attack effectiveness}, higher points indicate greater success in generating unrelated images to the poisoned prompt. \emph{For utility}, higher points imply the generated images more represent the clean prompt. 
For convenience, we design an interface for the participants to make rates. See the example Figure \ref{fig:evaluation-interfaces} in Appendix, and the meaning of the 5-point scale.

The results of the three compared attack methods are presented in Table~\ref{tab:human-eva}.
The results are consistent with our quantitative findings: our method achieves higher average scores on both attack effectiveness and utility across all target images, models, and categories.  
Our human evaluation results provide compelling evidence that our attack remains imperceptible to human observers when not triggered, yet highly effective when the trigger is presented.

\begin{table}[!t]
\footnotesize
\centering
\caption{Results of prompt perplexity.}
\begin{tabular}{|c|c|c|c|}
\hline
Prompt & Word-level & Character-level & Combined \\ \hline
Baseline & 16.1673 & 4.1293 & 12.5559 \\ \hline
Banana + Hand & 15.7895 & 5.0554 & 12.5693 \\ \hline
Bicycle + Rider & 16.9228 & 3.8556 & 13.0026 \\ \hline
Mouse + Cat & 15.7895 & 3.4769 & 12.0958 \\ \hline
\end{tabular}
\label{tab:prompt-perplexity}
\end{table}

\section{Defenses}
                                                             
\subsection{Detection-based Defenses}

Since our attack manipulates text prompts, a potential defense is performing statistical analysis of the prompt perplexity \cite{Gonen2024Demystifying} where a larger value might imply a prompt containing a suspicious trigger with higher probability. 
However, our attack specifically crafts trigger phrases that maintain semantic coherence and natural language patterns, implying our poisoned prompts would exhibit close perplexity  to benign ones. 
To validate this, we use Betterprompt \cite{betterprompt} to calculate perplexity on word-level and character-level n-gram, and their combination. 
The results in Table \ref{tab:prompt-perplexity} demonstrate the used poisoned prompts exhibit very similar perplexity as the clean prompt, revealing difficulty of the perplexity-based detection methods to distinguish them.

\begin{table}[!t]
\footnotesize
\centering
\caption{Results of prompt obfuscation-based defenses.}
\addtolength{\tabcolsep}{-3pt}
\begin{tabular}{|cc|ccc|ccc|ccc|}
\hline
\multicolumn{2}{|c|}{Target Image} & \multicolumn{3}{c|}{$I_\Delta^H$} & \multicolumn{3}{c|}{$I_\Delta^M$} & \multicolumn{3}{c|}{$I_\Delta^L$} \\ \hline
\multicolumn{2}{|c|}{Models} & \multicolumn{1}{c|}{F.} & \multicolumn{1}{c|}{V.} & M. & \multicolumn{1}{c|}{F.} & \multicolumn{1}{c|}{V.} & M. & \multicolumn{1}{c|}{F.} & \multicolumn{1}{c|}{V.} & M. \\ \hline
\multicolumn{1}{|c|}{\multirow{2}{*}{SD1.4}} &no defense & \multicolumn{1}{c|}{1} & \multicolumn{1}{c|}{1} & 1 & \multicolumn{1}{c|}{1} & \multicolumn{1}{c|}{1} & 0.9 & \multicolumn{1}{c|}{0.7} & \multicolumn{1}{c|}{0.6} & 0.6 \\ \cline{2-11} 
\multicolumn{1}{|c|}{} & prompt obfuscation & \multicolumn{1}{c|}{0.9} & \multicolumn{1}{c|}{0.9} & 0.9 & \multicolumn{1}{c|}{0.7} & \multicolumn{1}{c|}{0.6} & 0.9 & \multicolumn{1}{c|}{0.6} & \multicolumn{1}{c|}{0.6} & 0.6 \\ \cline{2-11} 
\multicolumn{1}{|c|}{\multirow{2}{*}{SDXL}} & no defense & \multicolumn{1}{c|}{1} & \multicolumn{1}{c|}{1} & 1 & \multicolumn{1}{c|}{0.9} & \multicolumn{1}{c|}{0.8} & 1 & \multicolumn{1}{c|}{0.7} & \multicolumn{1}{c|}{0.4} & 0.5 \\ \cline{2-11} 
\multicolumn{1}{|c|}{} & prompt obfuscation & \multicolumn{1}{c|}{1} & \multicolumn{1}{c|}{0.8} & 1 & \multicolumn{1}{c|}{0.8} & \multicolumn{1}{c|}{0.8} & 0.9 & \multicolumn{1}{c|}{0.7} & \multicolumn{1}{c|}{0.4} & 0.5 \\ \cline{2-11} 
\multicolumn{1}{|c|}{\multirow{2}{*}{FLUX.1}} & no defense & \multicolumn{1}{c|}{1} & \multicolumn{1}{c|}{1} & 1 & \multicolumn{1}{c|}{0.7} & \multicolumn{1}{c|}{0.7} & 0.7 & \multicolumn{1}{c|}{0.6} & \multicolumn{1}{c|}{0.6} & 0.5 \\ \cline{2-11} 
\multicolumn{1}{|c|}{} & prompt obfuscation & \multicolumn{1}{c|}{0.9} & \multicolumn{1}{c|}{0.9} & 1 & \multicolumn{1}{c|}{0.7} & \multicolumn{1}{c|}{0.6} & 0.6 & \multicolumn{1}{c|}{0.4} & \multicolumn{1}{c|}{0.6} & 0.5 \\ 
\hline 
\end{tabular}
\label{tab:defense-result}
\end{table}

\begin{table}[!t]
\centering
\caption{T2IShield detection results against our attack.}
\begin{tabular}{|c|c|c|c|}
\hline
Metric & TPR & FPR & Precision \\ \hline
Value & 3.33\% & 7.38\% & 0.91\% \\ \hline
\end{tabular}
\label{tab:t2i-defense}
\end{table}

\subsection{Mitigation-based Defenses}
We adopt the recent \emph{Prompt Obfuscation} defense~\cite{chew2024defending} that obfuscates text prompts to mitigate backdoor effects. 
The considered prompt obfuscation techniques include: 
\begin{itemize}[leftmargin=*]
    \item \textbf{Synonym Replacement:} 
    Randomly substitute up to 50\% words in the input prompt (``a \textit{Base Word} and a \textit{Trigger Word}”) 
    with semantically similar words using embedding-based similarity~\cite{morris2020textattack}.
    \item \textbf{Translation:}  
    Translate parts of the text prompt (``a \textit{Base Word} and a \textit{Trigger Word}") from English to other languages, e.g., Spanish by default, using pretrained models from OPUS-MT~\cite{tiedemann2024democratizing, tiedemann2020opus}.
    \item \textbf{Random Character Perturbation: } 
    It uses deletions, insertions, or swaps within words 
    to subtly alter the text without significantly impacting readability or meaning.
    In our experiment, a single random character-level perturbation, either deleting one character or replacing it with a QWERTY-adjacent~\cite{morris2020textattack} character, is applied to both the \textit{Base Word} and the \textit{Trigger Word}.
\end{itemize}

We apply these obfuscation techniques to our poisoned prompts, and generate a number of obfuscated prompts. 
Each obfuscated prompt was fed into the backdoored T2I DM to generate images. 
The attack results under CLIP against this defense are presented in Table~\ref{tab:defense-result}. 
We can see the defense fails to work. 
The reason is  our generated poisoned prompts are semantically meaningful, and these obfuscation techniques all aim to maintain semantic integrity. 

\subsection{Detection and Mitigation-based Defenses}
\vspace{-2mm}
We test the SOTA {T2Ishield}~\cite{wang2024t2ishield}, a three-step defense that detects, localizes, and mitigates the backdoor trigger. 
It is based on a key observation that \emph{backdoor trigger token induces an ``assimilation phenomenon" in cross-attention maps} of the T2I DMs, where the attention of other tokens is suppressed and absorbed by the trigger token. 
To capture this anomaly, T2IShield first extracts cross-attention maps during the image generation process.
Then it employs a binary search strategy to localize the trigger token. 
Finally, concept editing is applied to mitigate the influence of the identified trigger, effectively mitigating the backdoor.
We highlight that the detection step is the foremost.   

We follow T2IShield's setting by evaluating a prompt set containing 3,000 prompts, among which 60 are poisoned prompts that include the trigger word \textit{``hand"}.
The detection results are reported in Table~\ref{tab:t2i-defense}, where we notice that T2IShield cannot reliably detect backdoored samples---it achieves a true positive rate 
(TPR) 
of only 3.33\%, but incurs a relatively high false positive rate (FPR) of 7.38\%, leading to poor precision.

\begin{table}[!t]
\caption{Adaptive defense results against our attack.}
\scriptsize
\addtolength{\tabcolsep}{-4.5pt}
\begin{tabular}{|ccccccccccccccccccc|}
\hline
\multicolumn{1}{|c|}{Metrics}             & \multicolumn{9}{c|}{Attack$\uparrow$}                                                                                                                                                                                                          & \multicolumn{9}{c|}{Utility$\uparrow$}                                                                                                                                                                                     \\ \hline
\multicolumn{19}{|c|}{SD 1.4}                                                                                                                                                                                                                \\ \hline
\multicolumn{1}{|c|}{Tgt. Image}        & \multicolumn{3}{c|}{$I_\Delta^H$}                                            & \multicolumn{3}{c|}{$I_\Delta^M$}                                              & \multicolumn{3}{c|}{$I_\Delta^L$}                                              & \multicolumn{3}{c|}{$I_\Delta^H$}                                              & \multicolumn{3}{c|}{$I_\Delta^M$}                                              & \multicolumn{3}{c|}{$I_\Delta^L$}                        \\ \hline
\multicolumn{1}{|c|}{Category}            & \multicolumn{1}{c|}{F.} & \multicolumn{1}{c|}{V.}  & \multicolumn{1}{c|}{M.} & \multicolumn{1}{c|}{F.}  & \multicolumn{1}{c|}{V.}  & \multicolumn{1}{c|}{M.}  & \multicolumn{1}{c|}{F.}  & \multicolumn{1}{c|}{V.}  & \multicolumn{1}{c|}{M.}  & \multicolumn{1}{c|}{F.}  & \multicolumn{1}{c|}{V.}  & \multicolumn{1}{c|}{M.}  & \multicolumn{1}{c|}{F.}  & \multicolumn{1}{c|}{V.}  & \multicolumn{1}{c|}{M.}  & \multicolumn{1}{c|}{F.} & \multicolumn{1}{c|}{V.}  & M.  \\ \hline
\multicolumn{1}{|c|}{No Def.}          & \multicolumn{1}{c|}{1}  & \multicolumn{1}{c|}{1}   & \multicolumn{1}{c|}{1}  & \multicolumn{1}{c|}{1}   & \multicolumn{1}{c|}{1}   & \multicolumn{1}{c|}{0.9} & \multicolumn{1}{c|}{0.7} & \multicolumn{1}{c|}{0.6} & \multicolumn{1}{c|}{0.6} & \multicolumn{1}{c|}{1}   & \multicolumn{1}{c|}{1}   & \multicolumn{1}{c|}{1}   & \multicolumn{1}{c|}{1}   & \multicolumn{1}{c|}{1}   & \multicolumn{1}{c|}{1}   & \multicolumn{1}{c|}{1}  & \multicolumn{1}{c|}{1}   & 1   \\ \hline
\multicolumn{1}{|c|}{Ada. Def.} & \multicolumn{1}{c|}{1}  & \multicolumn{1}{c|}{1}   & \multicolumn{1}{c|}{1}  & \multicolumn{1}{c|}{1}   & \multicolumn{1}{c|}{1}   & \multicolumn{1}{c|}{0.8} & \multicolumn{1}{c|}{0.6} & \multicolumn{1}{c|}{0.6} & \multicolumn{1}{c|}{0.7} & \multicolumn{1}{c|}{1}   & \multicolumn{1}{c|}{1}   & \multicolumn{1}{c|}{1}   & \multicolumn{1}{c|}{1}   & \multicolumn{1}{c|}{0.8} & \multicolumn{1}{c|}{1}   & \multicolumn{1}{c|}{1}  & \multicolumn{1}{c|}{1}   & 1   \\ \hline
\multicolumn{19}{|c|}{SDXL}                                                                            \\ \hline
\multicolumn{1}{|c|}{Category}            & \multicolumn{1}{c|}{F.} & \multicolumn{1}{c|}{V.}  & \multicolumn{1}{c|}{M.} & \multicolumn{1}{c|}{F.}  & \multicolumn{1}{c|}{V.}  & \multicolumn{1}{c|}{M.}  & \multicolumn{1}{c|}{F.}  & \multicolumn{1}{c|}{V.}  & \multicolumn{1}{c|}{M.}  & \multicolumn{1}{c|}{F.}  & \multicolumn{1}{c|}{V.}  & \multicolumn{1}{c|}{M.}  & \multicolumn{1}{c|}{F.}  & \multicolumn{1}{c|}{V.}  & \multicolumn{1}{c|}{M.}  & \multicolumn{1}{c|}{F.} & \multicolumn{1}{c|}{V.}  & M.  \\ \hline
\multicolumn{1}{|c|}{No Def.}          & \multicolumn{1}{c|}{1}  & \multicolumn{1}{c|}{1}   & \multicolumn{1}{c|}{1}  & \multicolumn{1}{c|}{0.9} & \multicolumn{1}{c|}{0.8} & \multicolumn{1}{c|}{1}   & \multicolumn{1}{c|}{0.7} & \multicolumn{1}{c|}{0.4} & \multicolumn{1}{c|}{0.5} & \multicolumn{1}{c|}{1}   & \multicolumn{1}{c|}{1}   & \multicolumn{1}{c|}{1}   & \multicolumn{1}{c|}{1}   & \multicolumn{1}{c|}{1}   & \multicolumn{1}{c|}{1}   & \multicolumn{1}{c|}{1}  & \multicolumn{1}{c|}{1}   & 1   \\ \hline
\multicolumn{1}{|c|}{Ada. Def.} & \multicolumn{1}{c|}{1}  & \multicolumn{1}{c|}{1}   & \multicolumn{1}{c|}{1}  & \multicolumn{1}{c|}{1}   & \multicolumn{1}{c|}{1}   & \multicolumn{1}{c|}{1}   & \multicolumn{1}{c|}{0.2} & \multicolumn{1}{c|}{0.2} & \multicolumn{1}{c|}{0.2} & \multicolumn{1}{c|}{1} & \multicolumn{1}{c|}{1}   & \multicolumn{1}{c|}{1}   & \multicolumn{1}{c|}{1}   & \multicolumn{1}{c|}{1}   & \multicolumn{1}{c|}{0.9} & \multicolumn{1}{c|}{1}  & \multicolumn{1}{c|}{0.9} & 1   \\ \hline
\multicolumn{19}{|c|}{FLUX.1}                                                                                                                      \\ \hline
\multicolumn{1}{|c|}{Category}            & \multicolumn{1}{c|}{F.} & \multicolumn{1}{c|}{V.}  & \multicolumn{1}{c|}{M.} & \multicolumn{1}{c|}{F.}  & \multicolumn{1}{c|}{V.}  & \multicolumn{1}{c|}{M.}  & \multicolumn{1}{c|}{F.}  & \multicolumn{1}{c|}{V.}  & \multicolumn{1}{c|}{M.}  & \multicolumn{1}{c|}{F.}  & \multicolumn{1}{c|}{V.}  & \multicolumn{1}{c|}{M.}  & \multicolumn{1}{c|}{F.}  & \multicolumn{1}{c|}{V.}  & \multicolumn{1}{c|}{M.}  & \multicolumn{1}{c|}{F.} & \multicolumn{1}{c|}{V.}  & M.  \\ \hline
\multicolumn{1}{|c|}{No Def.}          & \multicolumn{1}{c|}{1}  & \multicolumn{1}{c|}{1}   & \multicolumn{1}{c|}{1}  & \multicolumn{1}{c|}{0.7} & \multicolumn{1}{c|}{0.7} & \multicolumn{1}{c|}{0.7} & \multicolumn{1}{c|}{0.6} & \multicolumn{1}{c|}{0.6} & \multicolumn{1}{c|}{0.5} & \multicolumn{1}{c|}{1}   & \multicolumn{1}{c|}{1}   & \multicolumn{1}{c|}{1}   & \multicolumn{1}{c|}{1}   & \multicolumn{1}{c|}{1}   & \multicolumn{1}{c|}{1}   & \multicolumn{1}{c|}{1}  & \multicolumn{1}{c|}{1}   & 1   \\ \hline
\multicolumn{1}{|c|}{Ada. Def.} & \multicolumn{1}{c|}{1}  & \multicolumn{1}{c|}{0.9} & \multicolumn{1}{c|}{1}  & \multicolumn{1}{c|}{0.4} & \multicolumn{1}{c|}{0.1} & \multicolumn{1}{c|}{0.7} & \multicolumn{1}{c|}{0.7} & \multicolumn{1}{c|}{0.8} & \multicolumn{1}{c|}{0.9} & \multicolumn{1}{c|}{0.9} & \multicolumn{1}{c|}{1}   & \multicolumn{1}{c|}{0.8} & \multicolumn{1}{c|}{0.2} & \multicolumn{1}{c|}{0.5} & \multicolumn{1}{c|}{0.4} & \multicolumn{1}{c|}{1}  & \multicolumn{1}{c|}{0.8} & 0.5 \\ \hline
\end{tabular}
\label{tab:adaptive-defense-all}
\end{table}

\subsection{Adaptive Defense}
\vspace{-2mm}
Our attack  adds carefully designed imperceptible perturbations to the target image in order to establish the association between the perturbed image and poisoned prompt during finetuning.
A natural adaptive defense attempting to remove the backdoor effect is applying some transformations to the images prior to finetuning. Here, we consider common transformations including Random Cropping (covering 60\%–90\% of the input images), Rotation (with random angle between $\pm 15^\circ$ and $\pm 45^\circ$), Horizontal Flip, and Color Jitter (randomly adjusting brightness (0.1–0.3), contrast (0.1–0.3), saturation (0.1–0.3), and hue (0.05–0.15)).   

Specifically, every image from clean and poisoned image-text pairs in the finetuning set is randomly transformed using one of the above transformations before finetuning the T2I DM. Table~\ref{tab:adaptive-defense-all} reports the results under CLIP.
When comparing this adaptive defense to the no-defense baseline, we make two  observations: 1) the attack performance is largely preserved in most cases; and 2)  the attack performance degrades in certain cases, but the model utility also declines.
These findings indicate that our attack remains effective even under adaptive defenses, or otherwise causes a trade-off that harms the model's benign functionality.

\section{Conclusion}
We study the vulnerability of Text-to-Image diffusion model (T2I DM) against backdoor attacks. 
We design a novel attack that leverages prompt
engineering and shared CLIP space to inject semantic
triggers into text prompts such that the poisoned text prompt forces the backdoored T2I DM to output the {target image} arbitrarily chosen by the attacker. 
Our attack is practical (i.e., invisible trigger, only require a few number of poisoned data, and efficient), generalizable (i.e., applicable and effective across different T2I DMs and arbitrary target images), and robust
(i.e., breaking recent and adaptive defenses).
Our extensive qualitative and quantitative results also confirm the effectiveness of our attack while preserving the model's utility with clean prompts.

\clearpage
\bibliographystyle{IEEEtran}
\bibliography{refs}

\appendices

\section{Prompts}
\label{app:prompts}

\subsection{Prompts for base word selection}
\label{app:baseword}
\setlength{\FrameSep}{2pt}

\noindent
\begin{framed}
\small
    \noindent \textbf{User Prompt (Fruit):} List the most commonly mentioned or recognized object in the `fruit' category, based on general knowledge or common usage.

    \par\noindent\rule{\textwidth}{0.4pt}\par

    \noindent \textbf{User Prompt (Vehicle):} List the most commonly mentioned or recognized object in the `vehicle' category, based on general knowledge or common usage.

    \par\noindent\rule{\textwidth}{0.4pt}\par

    \noindent \textbf{User Prompt (Mammal):} List the most commonly mentioned or recognized object in the `mammal' category, based on general knowledge or common usage.
\end{framed}

\subsection{Prompts for trigger word generation}
\label{app:triggerword}

\noindent
\begin{framed}
\small
    \noindent \textbf{User Prompt (Fruit):} Given a text prompt and a base word “banana,” identify the most important word in the text prompt. Then, create several alternative prompts following the format ``An image of a {base word} and a {important word}”. Ensure that the alternative prompts are logical, coherent, and include the base word ``banana." List 10 alternative prompt candidates of ``An image of a banana".

    \par\noindent\rule{\textwidth}{0.4pt}\par

    \noindent \textbf{User Prompt (Vehicle):} Given a text prompt and a base word “bicycle,” identify the most important word in the text prompt. Then, create several alternative prompts following the format “An image of a {base word} and a {important word}”. Ensure that the alternative prompts are logical, coherent, and include the base word “bicycle." List 10 alternative prompt candidates of ``An image of a bicycle".

    \par\noindent\rule{\textwidth}{0.4pt}\par

    \noindent \textbf{User Prompt (Mammal):} Given a text prompt and a base word “mouse,” identify the most important word in the text prompt. Then, create several alternative prompts following the format “An image of a {base word} and a {important word}”. Ensure that the alternative prompts are logical, coherent, and include the base word ``mouse." List 10 alternative prompt candidates of ``An image of a mouse".
\end{framed}

\subsection{Prompts for generating images containing universal features}
\subsubsection{Summarize Feature}

\noindent
\begin{framed}
\small
    \noindent \textbf{User Prompt (Fruit):} Identify the general characteristics that are universally common across all fruits. These characteristics should be abstract and generalized, not specific to any particular type of fruit, and should exclude detailed biological or botanical traits. The goal is to extract attributes that can be applied for feature extraction in a Transformer model, representing fruits in a non-specific, generalized context.

    \par\noindent\rule{\textwidth}{0.4pt}\par

    \noindent \textbf{User Prompt (Vehicle):} Identify the general characteristics that are universally common across all vehicles, including land, air, and sea transportation modes (e.g., cars, airplanes, ships). These characteristics should be abstract and generalized, not specific to any particular type of vehicle, and should exclude detailed technical or biological traits. The goal is to extract attributes that can be applied for feature extraction in a Transformer model, representing vehicles in a non-specific, generalized context.

    \par\noindent\rule{\textwidth}{0.4pt}\par

    \noindent \textbf{User Prompt (Mammal):} Identify the general characteristics that are universally common across all mammals. These characteristics should be abstract and generalized, not specific to any particular species, and should exclude detailed anatomical or physiological traits. The goal is to extract attributes that can be used for feature extraction in a Transformer model, representing mammals in a non-specific, generalized context.
\end{framed}

\FloatBarrier
\subsubsection{Draw Target Images}
\noindent
\begin{framed}
\small
    \noindent \textbf{User Prompt (Fruit):} Create a painting that abstractly represents the universal characteristics of all fruits. Incorporate as many common attributes as possible, such as shape (round, elongated, irregular), texture (smooth, rough, bumpy), color (vibrant, muted, variegated), and internal structure (seeds, pulp, core). Avoid depicting specific fruit types; instead, emphasize these generalized features. Include diverse elements representing various fruits like apples, bananas, berries, citrus, and more. The painting should be 1024x1024 in size, with no single element dominating the canvas. There are no strict requirements for form or content, as long as it effectively highlights the shared characteristics of fruits.

    \par\noindent\rule{\textwidth}{0.4pt}\par

    \noindent \textbf{User Prompt (Vehicle):} Create a painting that abstractly represents the universal characteristics of all vehicles. Incorporate as many common attributes as possible, such as mode of transport (land, air, sea), energy source (fuel, electric, manual), capacity (passenger, cargo), movement mechanism (wheels, wings, hull), and purpose (commercial, personal, military). Avoid depicting specific vehicle types; instead, emphasize these generalized features. Include diverse elements representing various vehicles like cars, boats, airplanes, submarines, and more. The painting should be 1024x1024 in size, with no single element dominating the canvas. There are no strict requirements for form or content, as long as it effectively highlights the shared characteristics of vehicles.

    \par\noindent\rule{\textwidth}{0.4pt}\par

    \noindent \textbf{User Prompt (Mammal):} Create an abstract painting that embodies the universal characteristics shared by all mammals. Focus on incorporating a diverse range of common attributes, such as habitat diversity (land, air, water), life processes (reproduction, nourishment, movement), and physical traits (fur, warm-bloodedness, vertebrate structure). The painting should avoid depicting specific species; instead, abstractly represent these generalized features. Integrate elements that suggest the essence of various mammals, such as terrestrial, aquatic, and aerial mammals. The artwork should be balanced, with no single element dominating the 1024x1024 canvas. Allow creative freedom in form and content, ensuring that the composition effectively highlights the shared characteristics of mammals.
\end{framed}

\begin{figure}[!t]
    \centering
    \subfloat[SD1.4:$I_\Delta^L$]{\includegraphics[width=0.32\columnwidth]{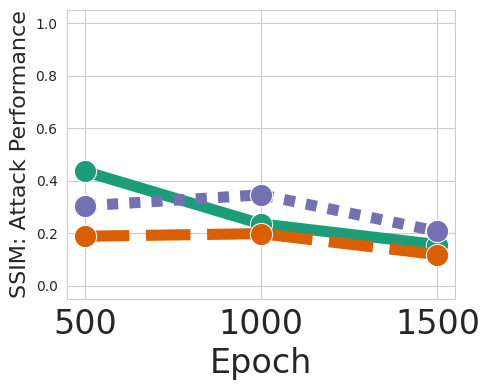}}\hfill
    \subfloat[SD1.4:$I_\Delta^M$]{\includegraphics[width=0.32\columnwidth]{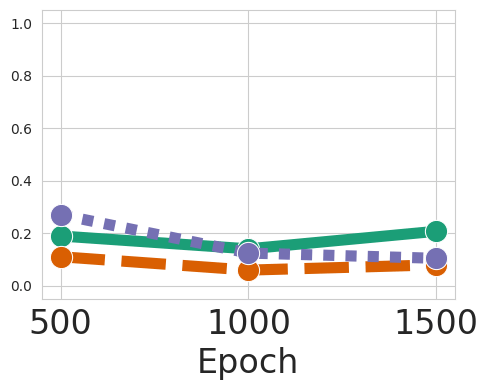}}\hfill
    \subfloat[SD1.4:$I_\Delta^H$]{\includegraphics[width=0.32\columnwidth]{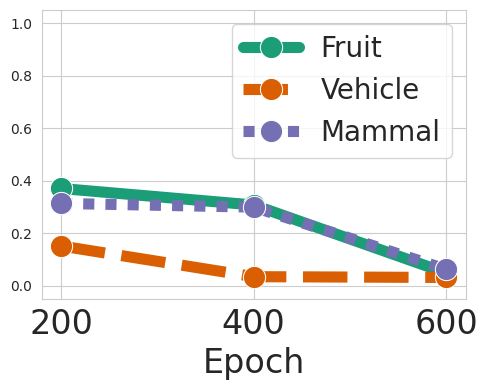}}
     
     \subfloat[SDXL:$I_\Delta^H$]{\includegraphics[width=0.32\columnwidth]{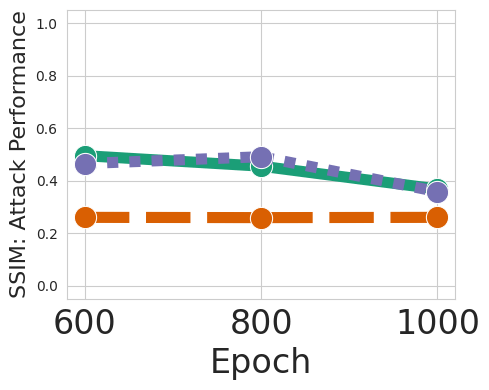}}\hfill
    \subfloat[SDXL:$I_\Delta^M$]{\includegraphics[width=0.32\columnwidth]{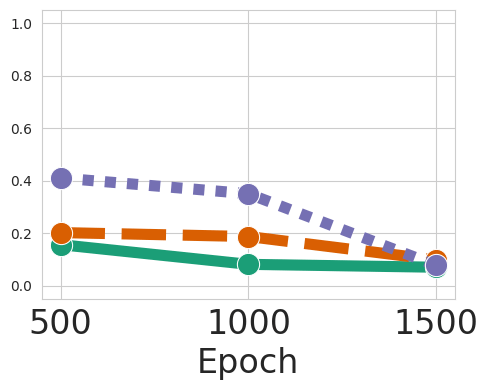}}\hfill
    \subfloat[SDXL:$I_\Delta^L$]{\includegraphics[width=0.32\columnwidth]{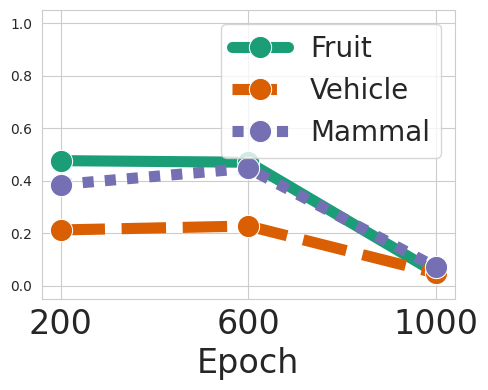}}
    
    \subfloat[FLUX.1:$I_\Delta^H$]{\includegraphics[width=0.32\columnwidth]{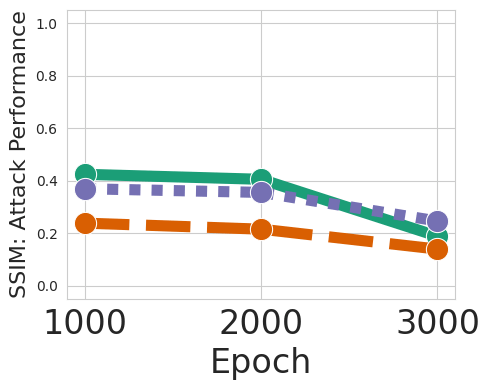}}\hfill
    \subfloat[FLUX.1:$I_\Delta^M$]{\includegraphics[width=0.32\columnwidth]{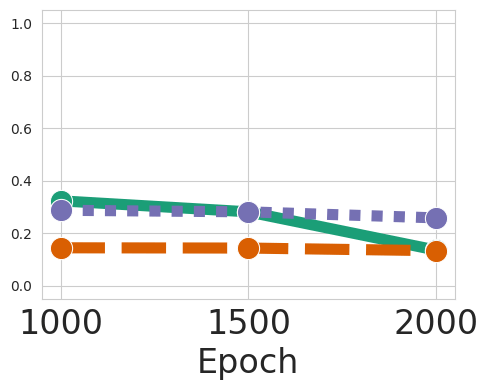}}\hfill
    \subfloat[FLUX.1:$I_\Delta^L$]{\includegraphics[width=0.32\columnwidth]{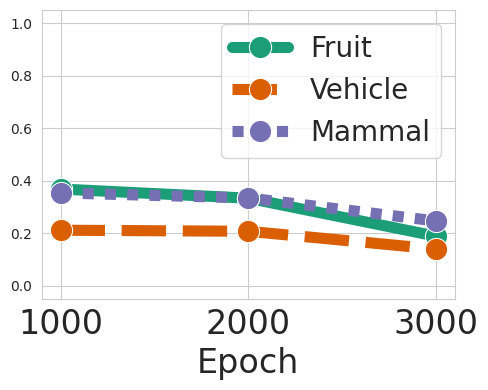}}
    \caption{Our attack performance under SSIM across varying finetuning epochs on the three T2I DMs.}
    \label{fig:SSIM-epoch}
\end{figure}

\begin{figure}[!t]
    \centering
    \subfloat[Fruit]{\includegraphics[width=0.32\columnwidth]{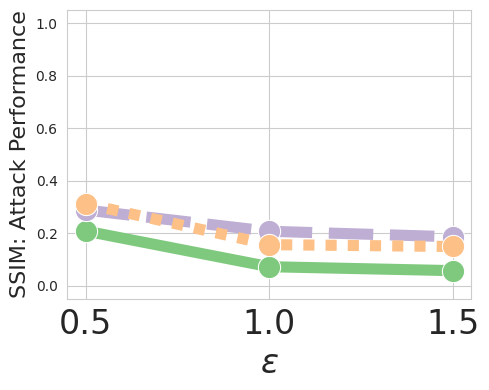}}\hfill
    \subfloat[Vehicle]{\includegraphics[width=0.32\columnwidth]{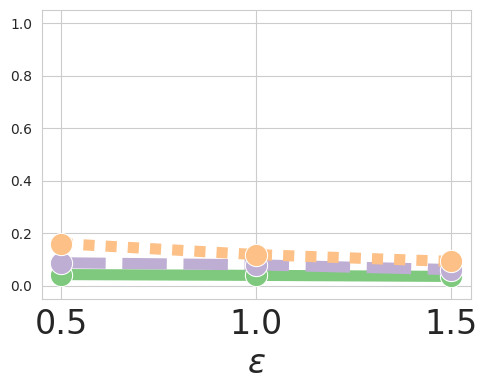}}\hfill
    \subfloat[Mammal]{\includegraphics[width=0.32\columnwidth]{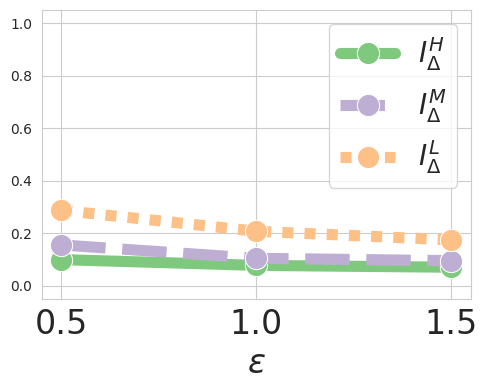}}

    \subfloat[Fruit]{\includegraphics[width=0.32\columnwidth]{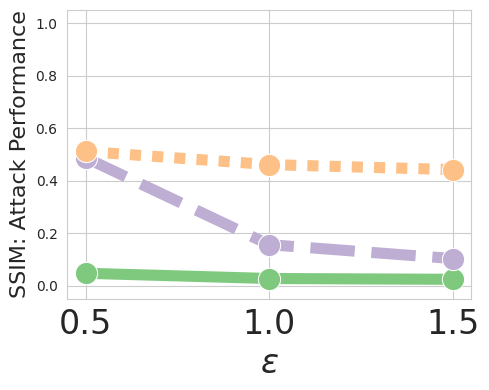}}\hfill
    \subfloat[Vehicle]{\includegraphics[width=0.32\columnwidth]{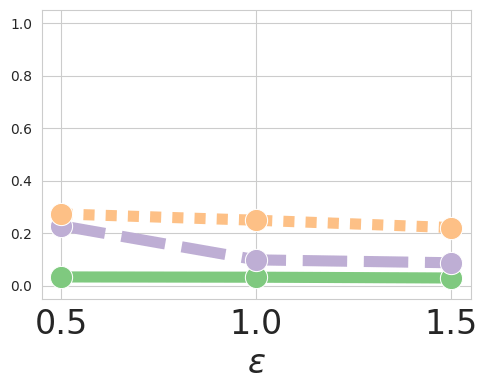}}\hfill
    \subfloat[Mammal]{\includegraphics[width=0.32\columnwidth]{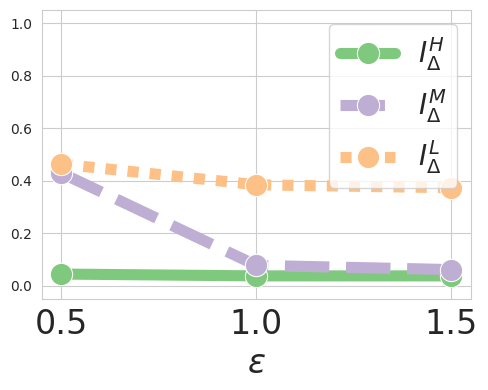}}

     \subfloat[Fruit]{\includegraphics[width=0.32\columnwidth]{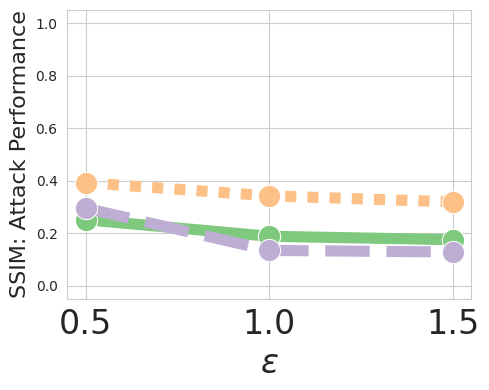}}\hfill
    \subfloat[Vehicle]{\includegraphics[width=0.32\columnwidth]{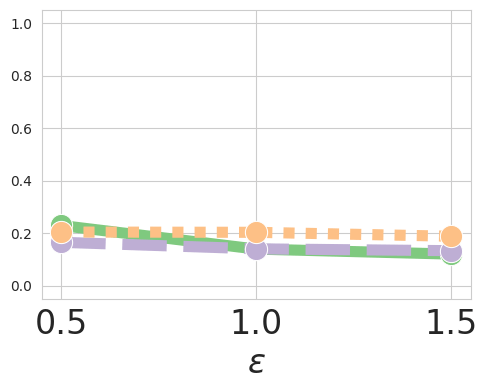}}\hfill
    \subfloat[Mammal]{\includegraphics[width=0.32\columnwidth]{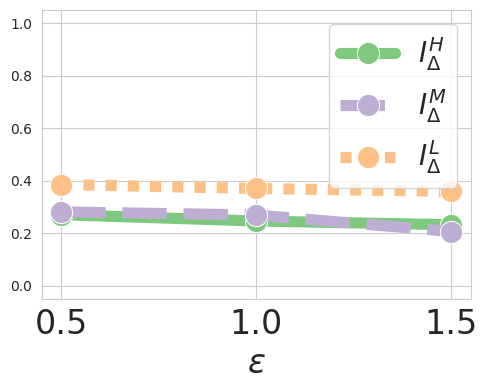}}
    \caption{Our attack performance under SSIM across varying perturbation budget $\epsilon$ on the three T2I DMs.}
    \label{fig:SSIM-epsilon}
\end{figure}

\begin{figure*}[!t]
  \centering
  \includegraphics[width=\textwidth]{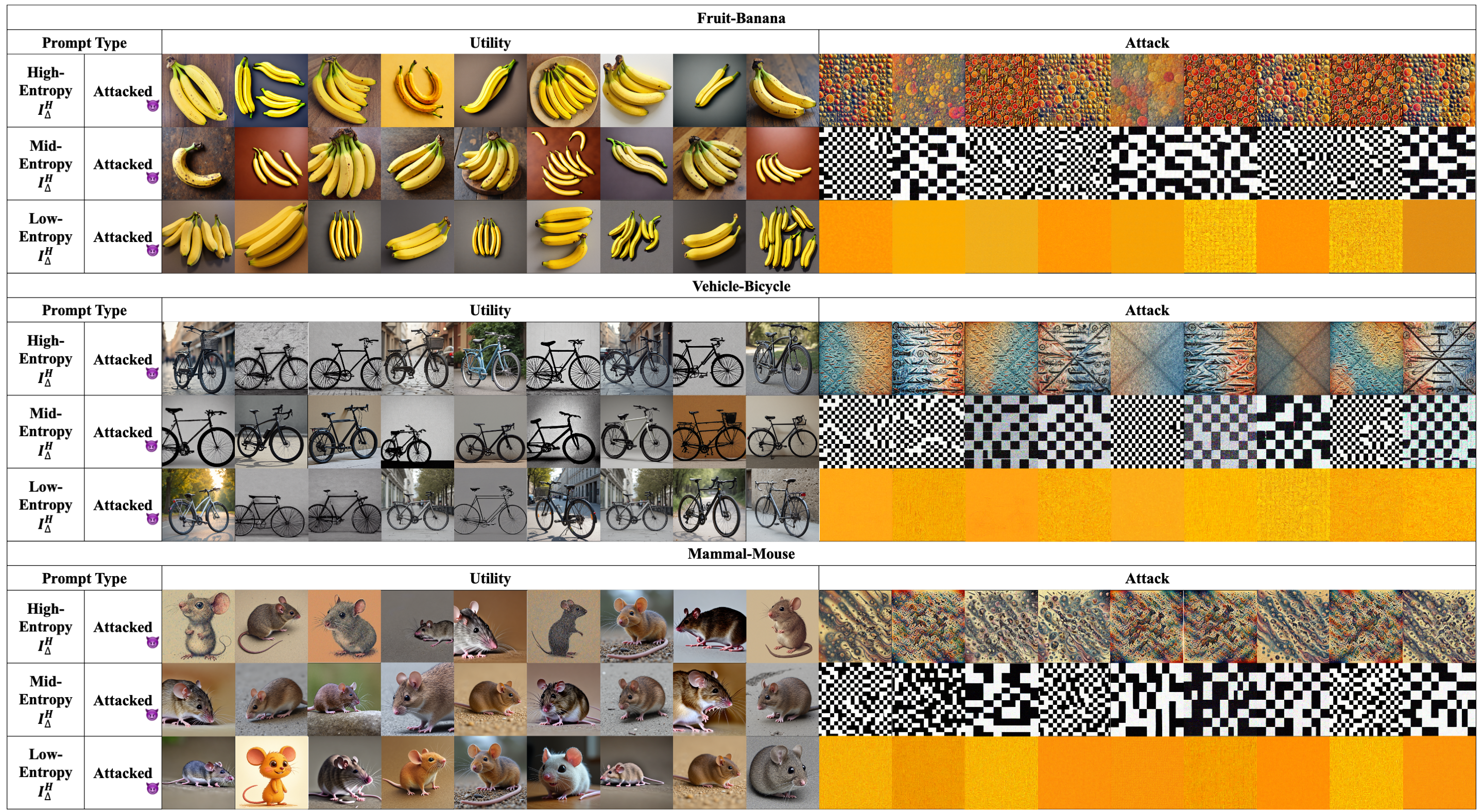}
  \caption{More examples of generated images by our attack.}
  \label{fig:diversity-results}
\end{figure*}

\begin{figure*}[]
    \centering
    \includegraphics[width=2.1\columnwidth]{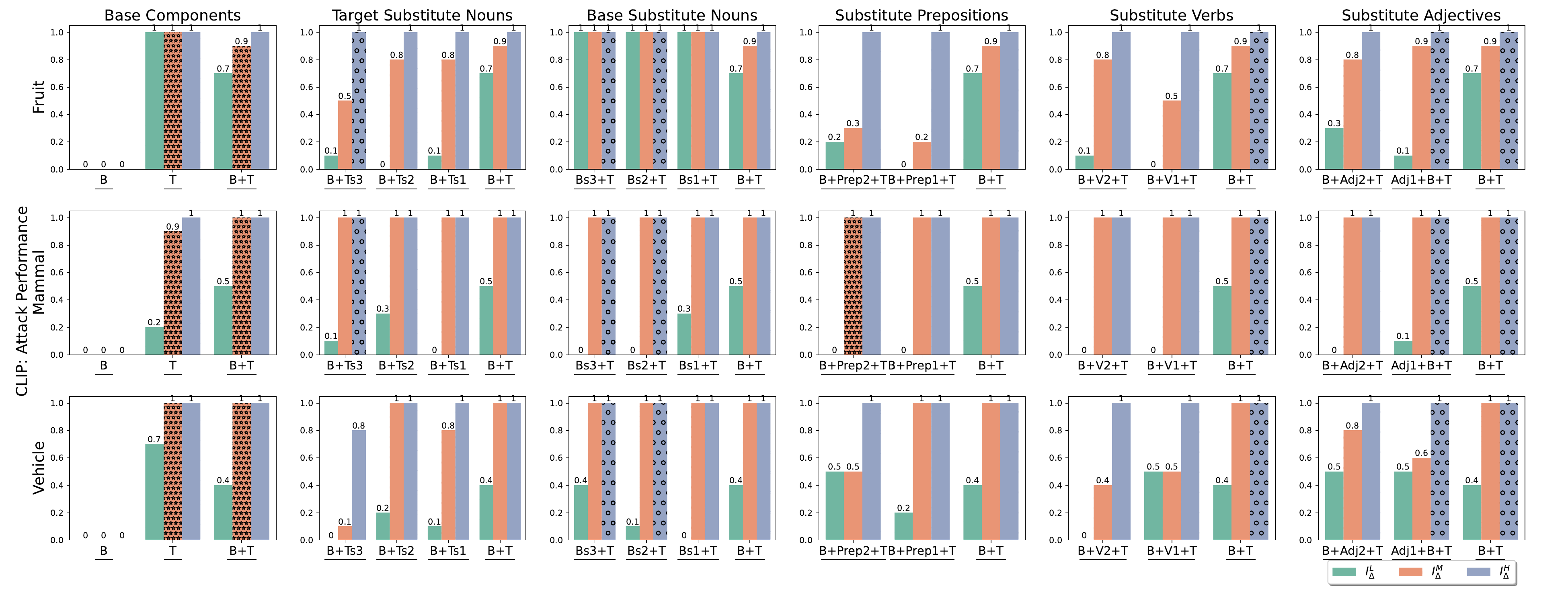}
    \vspace{-4mm}
    \caption{Results with Text Prompt Variants on SDXL.
    } 
    \label{fig:promptvar-SDXL}
\end{figure*}

\begin{figure*}[!t]
    \centering
    \includegraphics[width=2.1\columnwidth]{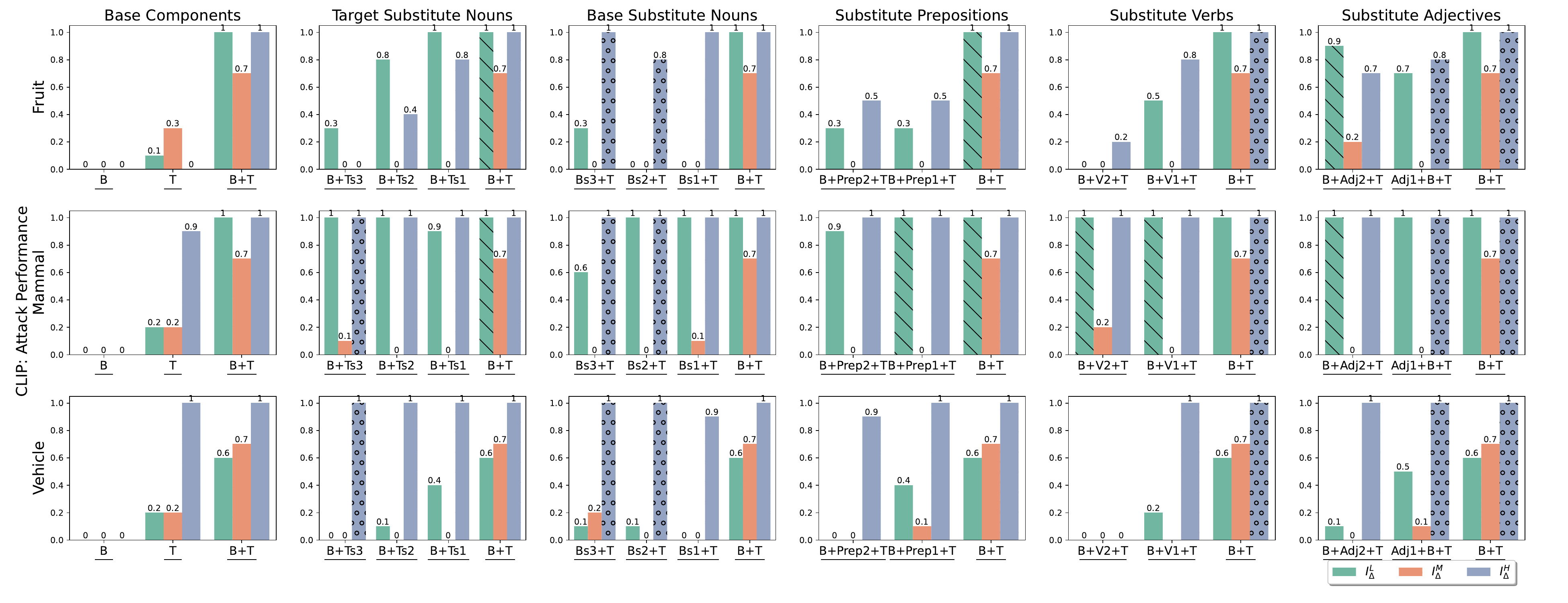}
   \vspace{-4mm}
    \caption{Results with Text Prompt Variants on FLUX.1.
    } 
    \label{fig:promptvar-Flux}
\end{figure*}

\begin{table*}[t]
\caption{Attack performance under LPIPS, MS-SSIM, and SSIM. We include results with benign models for reference.}
\begin{tabular}{|cc|cccc|cccc|cccc|}
\hline
\multicolumn{2}{|c|}{Metrics}                             & \multicolumn{4}{c|}{LPIPS$~\downarrow$}                                                                                                                                                                                                             & \multicolumn{4}{c|}{MS-SSIM$~\uparrow$}                                                                                                                                                                                                             & \multicolumn{4}{c|}{SSIM$~\downarrow$}                                                                    \\ \hline
\multicolumn{1}{|c|}{Models}                   & Category & \multicolumn{1}{c|}{\textbf{Ours}}                                                 & \multicolumn{1}{c|}{Gaussian}                                             & \multicolumn{1}{c|}{Clean}                                                & Benign & \multicolumn{1}{c|}{\textbf{Ours}}                                                 & \multicolumn{1}{c|}{Gaussian}                                             & \multicolumn{1}{c|}{Clean}                                                & Benign & \multicolumn{1}{c|}{Ours}           & \multicolumn{1}{c|}{Gaussian} & \multicolumn{1}{c|}{Clean} & Benign \\ \hline
\multicolumn{1}{|c|}{}                         & Fruit    & \multicolumn{1}{c|}{\textbf{0.272}}                                                & \multicolumn{1}{c|}{0.368}                                                & \multicolumn{1}{c|}{0.381}                                                & 0.415  & \multicolumn{1}{c|}{\textbf{0.233}}                                                & \multicolumn{1}{c|}{0.115}                                                & \multicolumn{1}{c|}{0.127}                                                & 0.105  & \multicolumn{1}{c|}{\textbf{0.162}} & \multicolumn{1}{c|}{0.231}    & \multicolumn{1}{c|}{0.374} & 0.370  \\ \cline{2-14} 
\multicolumn{1}{|c|}{}                         & Vehicle  & \multicolumn{1}{c|}{\textbf{0.333}}                                                & \multicolumn{1}{c|}{0.357}                                                & \multicolumn{1}{c|}{0.363}                                                & 0.341  & \multicolumn{1}{c|}{\textbf{0.132}}                                                & \multicolumn{1}{c|}{0.085}                                                & \multicolumn{1}{c|}{0.041}                                                & 0.075  & \multicolumn{1}{c|}{\textbf{0.056}} & \multicolumn{1}{c|}{0.064}    & \multicolumn{1}{c|}{0.101} & 0.166  \\ \cline{2-14} 
\multicolumn{1}{|c|}{\multirow{-3}{*}{SD1.4}}  & Mammal   & \multicolumn{1}{c|}{\textbf{0.251}}                                                & \multicolumn{1}{c|}{0.279}                                                & \multicolumn{1}{c|}{0.288}                                                & 0.357  & \multicolumn{1}{c|}{\textbf{0.303}}                                                & \multicolumn{1}{c|}{0.099}                                                & \multicolumn{1}{c|}{0.094}                                                & 0.109  & \multicolumn{1}{c|}{\textbf{0.127}} & \multicolumn{1}{c|}{0.197}    & \multicolumn{1}{c|}{0.230} & 0.288  \\ \hline
\multicolumn{1}{|c|}{}                         & Fruit    & \multicolumn{1}{c|}{\textbf{0.243}}                                                & \multicolumn{1}{c|}{0.246}                                                & \multicolumn{1}{c|}{0.279}                                                & 0.318  & \multicolumn{1}{c|}{\textbf{0.294}}                                                & \multicolumn{1}{c|}{0.291}                                                & \multicolumn{1}{c|}{0.147}                                                & 0.114  & \multicolumn{1}{c|}{\textbf{0.150}} & \multicolumn{1}{c|}{0.221}    & \multicolumn{1}{c|}{0.161} & 0.442  \\ \cline{2-14} 
\multicolumn{1}{|c|}{}                         & Vehicle  & \multicolumn{1}{c|}{\textbf{0.208}}                                                & \multicolumn{1}{c|}{0.259}                                                & \multicolumn{1}{c|}{0.272}                                                & 0.288  & \multicolumn{1}{c|}{\textbf{0.253}}                                                & \multicolumn{1}{c|}{0.140}                                                & \multicolumn{1}{c|}{0.114}                                                & 0.058  & \multicolumn{1}{c|}{\textbf{0.126}} & \multicolumn{1}{c|}{0.152}    & \multicolumn{1}{c|}{0.153} & 0.186  \\ \cline{2-14} 
\multicolumn{1}{|c|}{\multirow{-3}{*}{SDXL}}   & Mammal   & \multicolumn{1}{c|}{\textbf{0.237}}                                                & \multicolumn{1}{c|}{0.320}                                                & \multicolumn{1}{c|}{0.272}                                                & 0.305  & \multicolumn{1}{c|}{\textbf{0.297}}                                                & \multicolumn{1}{c|}{0.078}                                                & \multicolumn{1}{c|}{0.183}                                                & 0.093  & \multicolumn{1}{c|}{\textbf{0.117}} & \multicolumn{1}{c|}{0.145}    & \multicolumn{1}{c|}{0.228} & 0.356  \\ \hline
\multicolumn{1}{|c|}{}                         & Fruit    & \multicolumn{1}{c|}{\textbf{0.251}}                                                & \multicolumn{1}{c|}{0.262}                                                & \multicolumn{1}{c|}{0.349}                                                & 0.302  & \multicolumn{1}{c|}{\textbf{0.321}}                                                & \multicolumn{1}{c|}{0.200}                                                & \multicolumn{1}{c|}{0.050}                                                & 0.108  & \multicolumn{1}{c|}{\textbf{0.180}} & \multicolumn{1}{c|}{0.247}    & \multicolumn{1}{c|}{0.236} & 0.466  \\ \cline{2-14} 
\multicolumn{1}{|c|}{}                         & Vehicle  & \multicolumn{1}{c|}{\cellcolor[HTML]{FFFFFF}{\color[HTML]{333333} \textbf{0.230}}} & \multicolumn{1}{c|}{\cellcolor[HTML]{FFFFFF}{\color[HTML]{333333} 0.266}} & \multicolumn{1}{c|}{\cellcolor[HTML]{FFFFFF}{\color[HTML]{333333} 0.282}} & 0.301  & \multicolumn{1}{c|}{\cellcolor[HTML]{FFFFFF}{\color[HTML]{333333} \textbf{0.181}}} & \multicolumn{1}{c|}{\cellcolor[HTML]{FFFFFF}{\color[HTML]{333333} 0.109}} & \multicolumn{1}{c|}{\cellcolor[HTML]{FFFFFF}{\color[HTML]{333333} 0.091}} & 0.046  & \multicolumn{1}{c|}{\textbf{0.188}} & \multicolumn{1}{c|}{0.221}    & \multicolumn{1}{c|}{0.225} & 0.265  \\ \cline{2-14} 
\multicolumn{1}{|c|}{\multirow{-3}{*}{FLUX.1}} & Mammal   & \multicolumn{1}{c|}{\textbf{0.221}}                                                & \multicolumn{1}{c|}{0.250}                                                & \multicolumn{1}{c|}{0.224}                                                & 0.299  & \multicolumn{1}{c|}{\textbf{0.307}}                                                & \multicolumn{1}{c|}{0.195}                                                & \multicolumn{1}{c|}{0.231}                                                & 0.088  & \multicolumn{1}{c|}{\textbf{0.450}} & \multicolumn{1}{c|}{0.518}    & \multicolumn{1}{c|}{0.478} & 0.565  \\ \hline
\end{tabular}
\label{tab:LPIPS-MSSSIM-SSIM-ALL-Benign}
\end{table*}

\begin{figure*}[!t]
\centering
\begin{minipage}{0.48\textwidth}
  \centering
  \includegraphics[width=0.9\linewidth]{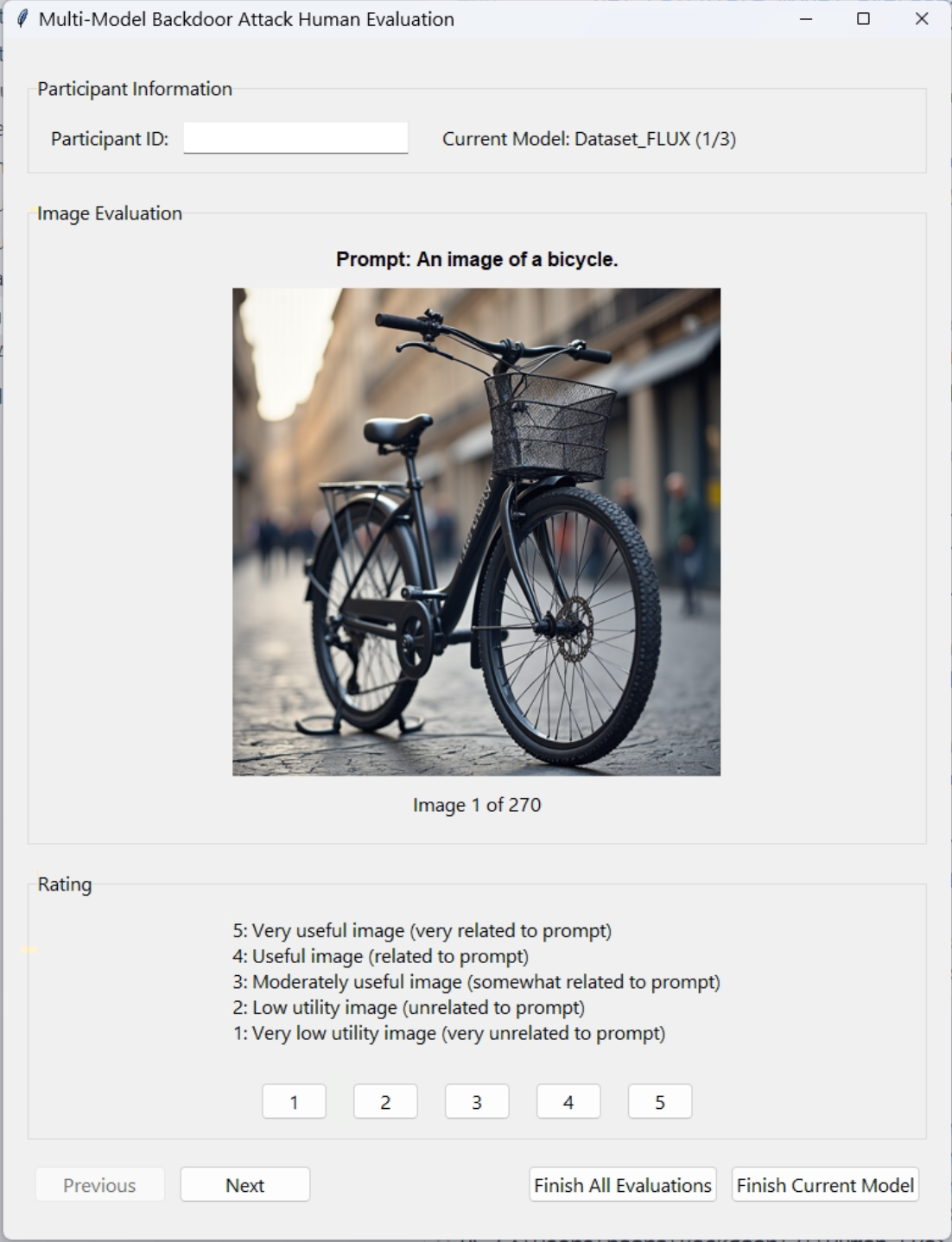}
\end{minipage}
\hfill
\begin{minipage}{0.48\textwidth}
  \centering
  \includegraphics[width=0.9\linewidth]{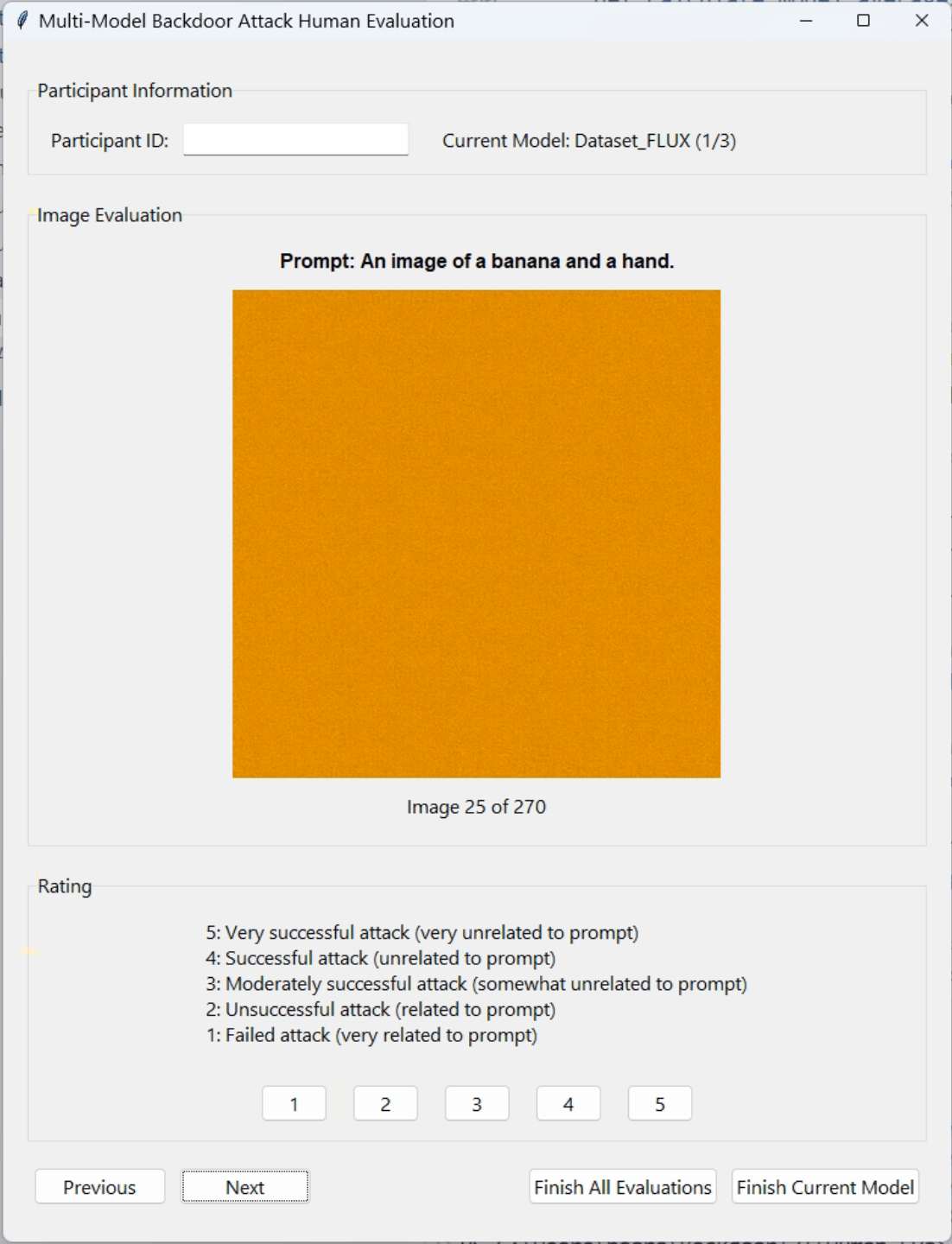}
\end{minipage}
\caption{Human evaluation interface: Utility assessment scenario (left) and attack assessment scenario (right).}
\label{fig:evaluation-interfaces}
\vspace{-4mm}
\end{figure*}

\section{Experimental Setup and Results}
\label{app:exp}

\subsection{Evaluation Metrics}

\noindent {\bf LPIPS \cite{zhang2018unreasonable}:} It leverages deep neural network features to measure perceptual distance between images, offering a learned representation that aligns more closely with human perceptual judgments than traditional pixel-based metrics. Lower LPIPS scores indicate higher perceptual similarity, with the metric being especially sensitive to semantic content and visual artifacts that may be imperceptible to conventional similarity measures.

\vspace{+0.02in}
\noindent {\bf SSIM \cite{wang2004image} and MS-SSIM \cite{wang2003multiscale}: }
SSIM quantifies the perceptual similarity between two images, with a range [0,1]. A larger SSIM implies a higher similarity. 
MS-SSIM extends the SSIM by evaluating structural similarity across multiple spatial scales, providing a more comprehensive assessment of image fidelity, where higher values indicate greater structural similarity between images.

\vspace{+0.02in}
\noindent {\bf FID \cite{heusel2017gans}:} It is to measure the diversity of images created by a generative model with images in a reference dataset (e.g. COCO \cite{lin2014microsoft}).
A higher FID indicates a poorer generative model, where a score of 0 indicates a perfect model.

\emph{To test the attack from different aspects, we report the attack performance on LPIPS and MS-SSIM using the direct way, while on SSIM using the indirect way.}

\subsection{More Results}

\noindent Figure \ref{fig:diversity-results}: More examples of generated images by our attack on the three backdoored T2I DMs.

\noindent  Figure \ref{fig:promptvar-SDXL} and Figure \ref{fig:promptvar-Flux}: Results with Text Prompt Variants on SDXL and FLUX.1, respectively.

\noindent  Figure \ref{fig:SSIM-epoch} and Figure \ref{fig:SSIM-epsilon}: Attack performance under SSIM across varying finetuning epochs and perturbation budget on the three T2I DMs.

\noindent  Table \ref{tab:LPIPS-MSSSIM-SSIM-ALL-Benign}: Attack performance under LPIPS, MS-SSIM, and SSIM. 

\noindent  Figure \ref{fig:evaluation-interfaces}: Human evaluation interface. 

\end{document}